\documentclass[journal]{IEEEtran}
\usepackage{amsmath}
\usepackage{graphicx}
\usepackage{makecell}
\usepackage[colorlinks, urlcolor=blue, linkcolor=blue, anchorcolor=blue, citecolor=blue]{hyperref}
\usepackage{bm}
\usepackage{amsfonts}
\usepackage{array}
\usepackage{threeparttable}
\usepackage{color}
\definecolor{R}{rgb}{1,0,0}

\begin{document}

\title{Recurrent Feature Propagation and \\Edge Skip-Connections for \\Automatic Abdominal Organ Segmentation}

\author{Zefan~Yang,
	Di~Lin,
	and~Yi~Wang
	\thanks{Z. Yang and Y. Wang are with the National-Regional Key Technology Engineering Laboratory for Medical Ultrasound, Guangdong Key Laboratory for Biomedical Measurements and Ultrasound Imaging, School of Biomedical Engineering, Health Science Center, Shenzhen University, Shenzhen, China, and also with the Medical UltraSound Image Computing (MUSIC) Lab, Shenzhen, China, and also with the Marshall Laboratory of Biomedical Engineering, Shenzhen University, Shenzhen, China (corresponding author: Yi Wang, onewang@szu.edu.cn).}
	\thanks{D. Lin is with the College of Intelligence and Computing, Tianjin University, Tianjin, China.}
}




\maketitle

\begin{abstract}
Automatic segmentation of abdominal organs in computed tomography (CT) images can support radiation therapy and image-guided surgery workflows. 
Developing of such automatic solutions remains challenging mainly owing to complex organ interactions and blurry boundaries in CT images. 
To address these issues, we focus on effective spatial context modeling and explicit edge segmentation priors.
Accordingly, we propose a 3D network with four main components trained end-to-end including shared encoder, edge detector, decoder with edge skip-connections (ESCs) and recurrent feature propagation head (RFP-Head).
To capture wide-range spatial dependencies, the RFP-Head propagates and harvests local features through directed acyclic graphs (DAGs) formulated with recurrent connections in an efficient slice-wise manner, with regard to spatial arrangement of image units.
To leverage edge information, the edge detector learns edge prior knowledge specifically tuned for semantic segmentation by exploiting intermediate features from the encoder with the edge supervision.
The ESCs then aggregate the edge knowledge with multi-level decoder features to learn a hierarchy of discriminative features explicitly modeling complementarity between organs' interiors and edges for segmentation.
We conduct extensive experiments on two challenging abdominal CT datasets with eight annotated organs.
Experimental results show that the proposed network outperforms several state-of-the-art models, especially for the segmentation of small and complicated structures (gallbladder, esophagus, stomach, pancreas and duodenum).
\textit{The code will be publicly available}.
\end{abstract}


\IEEEpeerreviewmaketitle

\section{Introduction}
\IEEEPARstart{M}{ulti-organ} segmentation from medical images can support multiple clinical applications, such as computer-aided diagnosis (CAD), treatment planning and treatment delivery~\cite{van2011computer}.
In CAD, organ segmentations allow quantification of clinical parameters related to organ volume and shape, to support clinical decision-making.
Besides, accurate segmentations of target volumes and organs at risk (OARs) are central to planning radiotherapy (e.g., stereotactic body radiotherapy), of which the success highly depends on the control of radiation exposure to the targets and OARs~\cite{tang2019clinically}.

\begin{figure}[t]
	\centering
	\includegraphics[width=3.5in]{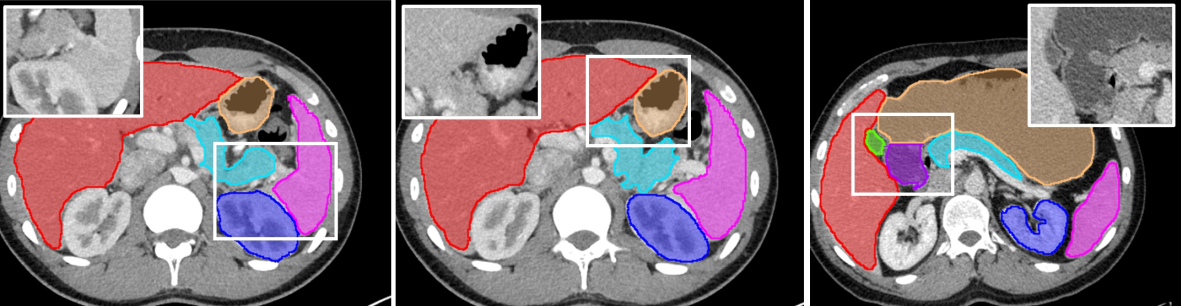}
	\caption{Representative abdominal CT images. Zoom-in image patches on top show structures of neighboring organs. The colored regions are annotated segmentations: spleen (magenta), left kidney (blue), gallbladder (green), liver (red), stomach (orange), pancreas (cyan) and duodenum (purple).}
	\label{fig1}
\end{figure}

Manual delineation of abdominal organs in computed tomography (CT) is time-consuming and laborious. 
Automatic multi-organ segmentation is a compelling solution but has two primary challenges. 
(1) Complicated internal structures:
Abdominal CT images contain several organs, such as liver, pancreas, stomach and duodenum. 
These organs have considerable shape, volume, appearance and position variations between and within patients, due to natural variability, soft tissue deformation, disease status and previous treatments. 
Moreover, since these organs huddle together in the rib cage, there exists complex interactions between spatially adjacent organs.
These factors result in complicated overall structures in abdominal CT volume.
Some representative images are displayed in Fig.~\ref{fig1}. 
(2) Low tissue contrast: Due to low-contrast acquisition of CT and small differences of attenuation coefficients in soft tissues, abdominal organs are close in values of Hounsfield unit (HU) (i.e., voxel intensity). 
Consequently, organs have weak boundaries between neighboring areas, for example between the head of duodenum and the stomach (see the zoom-in patch of the rightmost sample in Fig.~\ref{fig1}).


In this study, we propose a deep-learning-based algorithm for organ segmentation in CT images. 
In order to address aforementioned challenges, we suggest that effective spatial context modeling, as well as explicit edge priors, are demanding. 
Intuitively, when local appearance of the pancreas is not clearly distinguishable, a clinician normally recognizes and locates the pancreas based on surrounding anatomical landmarks (e.g., stomach, duodenum and liver). 
Inspired by this, our method aims to improve the representative capacity of local features by enabling the network to capture wide-range dependencies in a multi-dimensional recurrent neural network (MD-RNN)~\cite{graves2007multi} fashion.
In addition, medical images are typically acquired in standard anatomically aligned views with relatively structured organ shapes and orientations, 
thus we introduce explicit edge priors to provide extra constraint guidance, helping overcome the adverse effect of low soft tissue contrast.

\subsection{Related Work}
In the following, we review the literatures related to abdominal organ segmentation, context aggregation and edge detection, which are the main contributions of our new architecture.
\subsubsection{Abdominal Organ Segmentation}
Two branches of studies have been dedicated to abdominal organ segmentation.
One branch focused on shallow segmentation techniques, e.g., statistical shape models~\cite{zhang2010automatic,cerrolaza2015automatic} and multi-atlas label fusion~\cite{tong2015discriminative, xu2015efficient, shimizu2007segmentation}. 
These methods typically involve an image registration procedure.
Namely, the shape model-based methods use registration to estimate anatomical correspondences, while the atlas-based methods employ registration to transfer atlas images to the new image.
Consequently, their segmentation performance, especially for small organs, is mainly limited by the registration accuracy.

Another branch proposed end-to-end deep-learning frameworks. 
Fully convolutional network (FCN)~\cite{long2015fully} is an efficient architecture that enables pixel-wise prediction in one forward pass on natural images. 
Since its invention, many variants have been proposed for medical image segmentation.
One popular example is the 2D U-Net~\cite{ronneberger2015u}, the encoder-decoder architecture with skip-connections. 
For volumetric medical image segmentation, 3D architectures are more applicable.
There are a number of established architectures, including 3D UNet~\cite{cciccek20163d}, VNet~\cite{milletari2016v}, Attention UNet~\cite{oktay2018attention} and UNet++~\cite{zhou2019unet++}.

For abdominal organ segmentation, one group of studies focused on single organ segmentation, such as liver and tumor segmentation~\cite{li2018h,seo2019modified}, and pancreas segmentation~\cite{roth2018spatial,xue2019cascaded}.
In contrast, multi-organ segmentation poses more severe challenges due to complicated overall structures and complex organ-organ interactions.
In~\cite{gibson2017towards, gibson2018automatic}, Gibson \textit{et al}. incorporated dilated convolutions and dense skip-connections~\cite{huang2017densely} to segment abdominal organs.
Some other studies proposed two-stage cascaded networks, by either using coarse-to-fine segmentation strategies~\cite{roth2018multi,roth2018application,wang2019abdominal}, or focusing on the feature reuse~\cite{zhang2020block}.
Though these cascaded models offered advantages, they entailed high memory and computational overhead.
Also, it is effort-demanding to tune the optimal architectures.
To obtain an efficient model, Heinrich \textit{et al}.~\cite{heinrich2019obelisk} used sparse deformable convolutions to capture large spatial context. 




\subsubsection{Context Aggregation}
One branch of work used conditional random fields (CRFs) to model contextual dependencies of local regions.
The CRF is typically applied over class likelihood maps produced by FCNs.
It encourages pixels which are nearby in position/intensity to be assigned the same semantic label.
In~\cite{kamnitsas2017efficient}, Kamnitsas \textit{et al}. first extended the fully-connected CRF model~\cite{krahenbuhl2011efficient} to 3D segmentation.
Following that, several studies have employed the 3D fully-connected CRF for medical image segmentation and observed performance gains~\cite{christ2016automatic,alansary2016fast}.
However, the bilateral filter in the fully-connected CRF fails to model high-order contextual dependencies as it only constrains spatial and appearance consistency over local regions.

Other studies employed RNNs to model inherent contextual dependencies.
Poudel \textit{et al}.~\cite{poudel2016recurrent} recurrently integrated gated recurrent units (GRUs) into a 2D UNet architecture to segment left-ventricle in magnetic resonance images. 
Zhang \textit{et al}.~\cite{chen2016combining} combined bi-directional long short-term memory (LSTM) network with 2D UNet-like-architectures to conduct 3D segmentation. 
These studies used chain structured RNNs (i.e., 1D-RNNs) to exploit inter-slice context.
To capture rich spatial dependencies over 2D image regions, chain structured RNNs have been extended to 2D-RNNs.
Xie \textit{et al}.~\cite{xie2016spatial} proposed a spatial clockwork RNN to model contextual dependencies over image patches for perimysium segmentation. 
Shuai \textit{et al}.~\cite{shuai2017scene} placed a 2D-RNNs on top of a pretrained 2D FCN for scene segmentation.
Medical images commonly encompass a third dimension, but 2D architectures are unable to enforce inter-slice consistency.
Some previous work have explored the potentials of 3D-RNNs in medical image segmentation~\cite{stollenga2015parallel,andermatt2016multi}. 
However, due to extremely high computational costs, the 3D-RNNs are implemented on sub-volumes, thus have limited receptive fields.




\subsubsection{Edge Detection}
Studies have investigated different ways to incorporate boundary information into FCNs.
Several work designed loss functions to explicitly penalize errors on the segmentation boundary .
Anas \textit{et al}.~\cite{anas2017clinical} used an exponential weighted loss assigning higher values to pixels adjacent to boundary.
Karimi \textit{et al}.~\cite{karimi2019reducing} proposed loss functions derived from methods for estimating Hausdorff distance (HD), with the goal of directly reducing HD. 
More recently, Ma \textit{et al}.~\cite{ma2020learning} formulated the segmentation problem as a regression task of level set function, and used geodesic active contour as a supervision signal.
However, these methods are not directly applicable to multi-organ segmentation due to expensive computational costs incurred by distance transform, and also semantic ambiguity between adjacent areas of different classes.

Another group of work introduced sub-networks that predict edge maps by exploiting intermediate layer features. 
In~\cite{fan2020inf, zhang2019net}, the authors used low-level features to learn edge representations that were then combined with subsequent layer features using attention gates.
Zhou \textit{et al}.~\cite{zhou2019high} proposed to predict edge maps at the end of the network, which serves as auxiliary guidance for semantic segmentation.
These studies aim to explicitly learn edge knowledge to optimize segmentation performance.



\begin{figure}[t]
	\centering
	\includegraphics[width=3.5in]{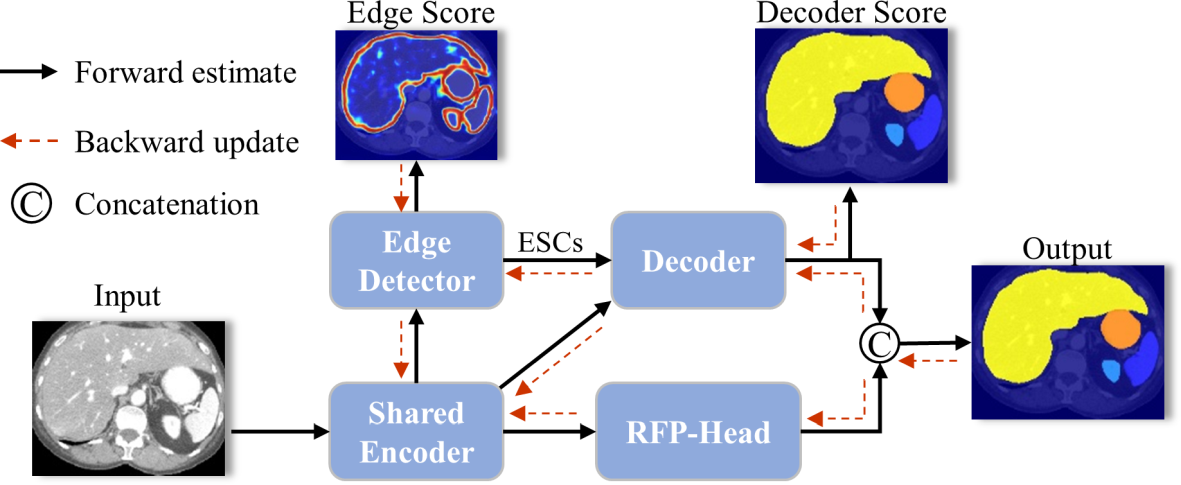}
	\caption{Overall framework of our proposed network. The network consists of four components: shared encoder, edge detector, decoder with edge skip-connections (ESCs) and RFP-Head. The output is predicted using complementary segmentation scores from the decoder and the RFP-Head. The full architecture is trained end-to-end by back-propagation (red dashed stream). }
	\label{fig2}
\end{figure}

\subsection{Contributions}
In this paper, we propose a network architecture with four main components: shared encoder, edge detector, decoder with edge skip-connections (ESCs), and recurrent feature propagation head (RFP-Head), as illustrated in Fig.~\ref{fig2}. 
The full architecture is trained end-to-end by back-propagation with volumetric CT images.
The shared encoder successively aggregates semantic information.
The designed RFP-Head is placed on top of the encoder to capture wide-range contextual dependencies over image regions by propagating and harvesting local features through directed graphs formulated with recurrent connections.
The RFP-Head processes the encoder output in an efficient manner and receives direct back-propagation gradients from the output loss function enabling effective optimization and fast convergence.
On the other hand,
the edge detector learns explicit edge priors tuned for semantic segmentation by exploiting the fine-grained and spatial invariant features from the shared encoder with edge supervision. 
The learned edge knowledge is then aggregated with multi-level decoder features by ESCs, allowing the intermediate layers to learn a hierarchy of discriminative features explicitly modeling complementarity between organs' interiors and edges.
Deep supervision~\cite{lee2015deeply} is used to aggregate multi-level discriminative features for segmentation predictions.

The main contributions of this paper can be summarized as follows:
\begin{itemize}
	\item We incorporate wide-range contextual dependencies to 3D FCNs by recurrently propagating and harvesting local features through directed graphs in an efficient slice-wise manner, in respect of spatial arrangement of image elements.
	
	\item We learn a task-specific edge detector tuned for semantic segmentation in an end-to-end trainable network. Edge skip-connections aggregate the learned edge priors with multi-level features enabling sufficient modeling of complementary information. 
	
	\item We conduct extensive experiments on two challenging multi-organ segmentation datasets. Experimental results show that our method outperforms several cutting-edge models, especially for the segmentation of small and complicated structures. 
\end{itemize}

\begin{figure*}[t]
	\centering
	\includegraphics[width=7.2in]{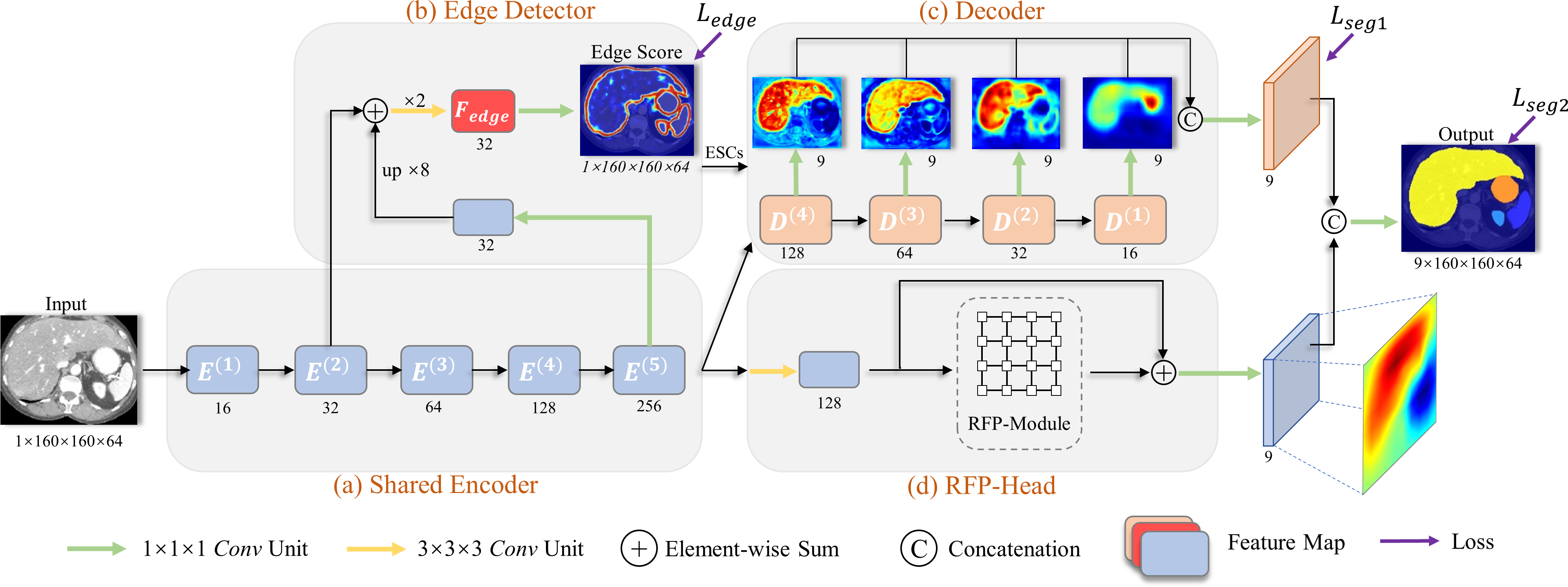}
	\caption{The proposed network architecture. First, the shared encoder computes five feature maps with 16, 32, 64, 128 and 256 channels using strided convolutions and max-pooling operation. Second, the edge detector leverages both the 2$^{nd}$ and the 5$^{th}$ stage features to learn an edge representation. Third, the decoder gradually upsamples the features to the input resolution, where edge skip-connections (ESCs) propagate the edge representation to each decoder block. Fourth, the RFP-Head recurrently propagates feature vectors via directed acyclic graphs (DAGs) to compute hidden feature vectors. Finally, the likelihood logits generated by both the decoder and the RFP-Head are concatenated to output the segmentation logits.}
	\label{fig3}
\end{figure*}

\section{Method}
As shown in Fig.~\ref{fig3}, the proposed network consists of four parts: (a) shared encoder, (b) edge detector, (c) decoder, and (d) RFP-Head. 
They are trained end-to-end to optimize the semantic segmentation performance.

\subsection{Encoder-Decoder Architecture}
We use the encoder-decoder architecture as backbone for multi-organ segmentation.
Each convolutional unit in our network comprises three functions: (1) a 3D convolution with a learned kernel, (2) a batch normalization layer, and (3) a rectified linear unit (ReLU) to enforce non-linearity.
The encoder uses consecutive strided convolutional unit to compute hierarchical features, and the decoder gradually recovering the feature maps to the input resolution.
We denote the encoder feature set as $E=\{E^{(1)}, E^{(2)}, E^{(3)}, E^{(4)}, E^{(5)}\}$, and the decoder feature set as $D=\{D^{(1)}, D^{(2)}, D^{(3)}, D^{(4)}\}$ (refer to Fig.~\ref{fig3} (a) and (c)).

The segmentation loss of the proposed network combines the cross-entropy loss and Dice loss~\cite{milletari2016v}, which is experimentally shown to be less sensitive to class imbalance:
\begin{equation}
	\begin{split}
		\mathcal{L}_{Dice} &= 1 - \sum_{k}^{K} \sum_{i}^{N} \frac{2 y_{k,i} \tilde{y}_{k,i}} {{y_{k,i}^{2}} + {\tilde{y}_{k,i}^{2}}}, \\
		\mathcal{L}_{CE} &= -\sum_{k}^{K} \sum_{i}^{N} y_{k,i} \log{(\tilde{y}_{k,i})}, \\
		\mathcal{L}_{seg} &= \mathcal{L}_{Dice} + \mathcal{L}_{CE},
	\end{split}
\end{equation}
where $K$ denotes the number of classes; $N$ denotes the total number of voxels. $y_{k,i}$ and $\tilde{y}_{k,i}$ denotes the ground-truth and predicted probability of class $k$ at voxel $i$, respectively.
The segmentation loss is imposed on both the decoder scores and final segmentation scores.

\subsection{Recurrent Feature Propagation}
In this section, we first elaborate the design of RFP-Module, the core building block of the RFP-Head, which consists of directed acyclic graphs (DAGs) that define the rules of feature propagation (see Fig.~\ref{fig4}), and recurrent connections for computing hidden features (see Fig.~\ref{fig5}). Then, we describe the workflow of our RFP-Head (see Fig.~\ref{fig3} (d)). 

\subsubsection{Directed Acyclic Graphs}
Some studies have applied chain-structured graph to structurize feature maps by sweeping across rows and columns~\cite{chen2016semantic,visin2016reseg}.
However, the connectivity structure of image units is beyond chain.
In other words, chain-structured graph loses spatial arrangement of image units, as the adjacent local features in image plane are not obligated to be neighbors in chain.
In this study, we adopt undirected cyclic graphs (UCGs) to represent the connectivity structure of image units.

\begin{figure}[t]
	\centering
	\includegraphics[width=3.5in]{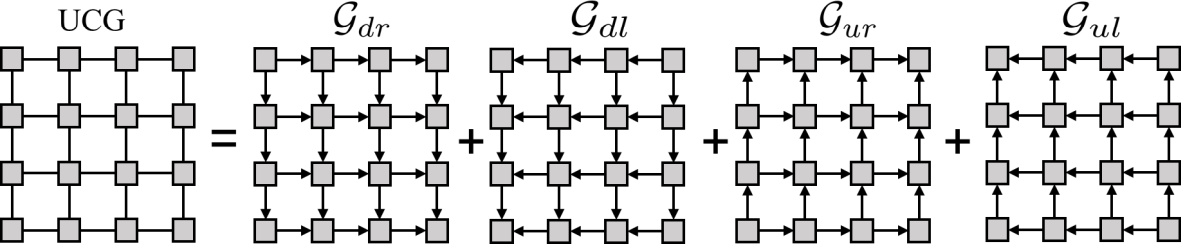}
	\caption{Decomposition of the undirected cyclic graph (UCG). The UCG is approximated by the combination of four directed acyclic graphs (DAGs), namely $\mathcal{G}_{dr}$, $\mathcal{G}_{dl}$, $\mathcal{G}_{ur}$ and $\mathcal{G}_{ul}$. For instance, the $\mathcal{G}_{dr}$ starts from top-left and scans down and right.}
	\label{fig4}
\end{figure}

\begin{figure}[t]
	\centering
	\includegraphics[width=3.5in]{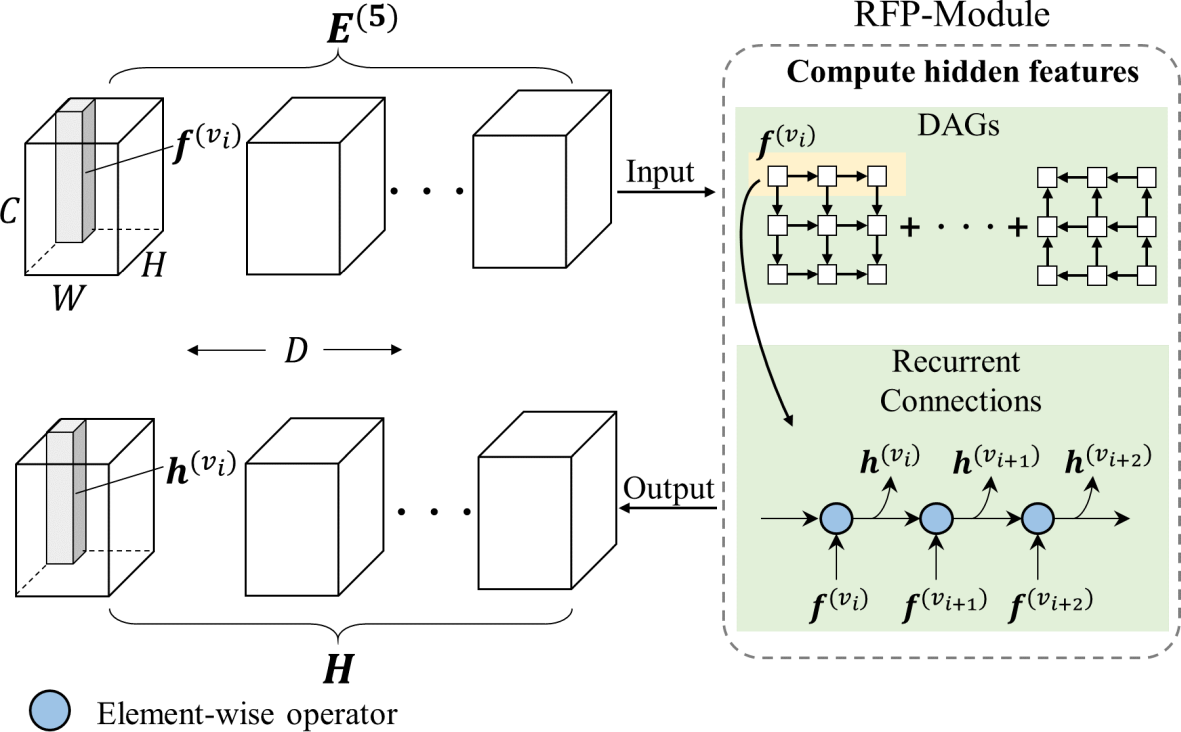}
	\caption{Flowchart of the proposed RFP-Module. The RFP-Module computes hidden features that capture wide-range dependencies in an efficient manner. First, the feature $\bm{E}^{(5)}$ is split into $D$ feature maps, which are then input to the RFP-Module. Second, the RFP-Module propagates the feature vectors $\bm{f}^{(v_i)}$ via directed acyclic graphs (DAGs) with recurrent connections, to compute hidden feature vectors $\bm{h}^{(v_i)}$. Finally, all hidden feature maps are concatenated to constitute the feature $\bm{H}$. For clarity, the recurrent connections are shown in 1D here.}
	\label{fig5}
\end{figure}

Due to the loopy property of UCGs, they are unable to be unfolded into acyclic computational graphs. Therefore, recurrent computation is not directly applicable to UCG-structured images.
To address this issue, we approximate the topology of UCGs by a combination of four layers of directed acyclic graphs (DAGs).
Namely, a UCG-structured image is represented as the combination of four DAG-structured images.
We denote four layers of DAGs sweeping from four directions as $\mathcal{G}_{dr}$, $\mathcal{G}_{dl}$, $\mathcal{G}_{ur}$, and $\mathcal{G}_{ul}$ (see Fig.~\ref{fig4}). For example, $\mathcal{G}_{dr}$ starts in the top-left vertex and scans down and right; $\mathcal{G}_{ul}$ starts in the bottom-right vertex and scans up and left, etc. All the layers are then added together to represent the connectivity of local neighborhood.


\subsubsection{Recurrent Connections}
The design of recurrent connections is applied in recurrent neural networks (RNNs) to equip the network with memory.
In this paper, we formulate the DAG-structured images with recurrent connections to perform context aggregation.
Specifically, as shown in Fig.~\ref{fig5}, we denote $v_{i}$ as the vertex at position $i$. We denote feature vector of $v_{i}$ as $\bm{f}^{(v_{i})}$, and hidden activation as $\bm{h}^{(v_{i})}$.
The forward propagation of DAG-structured image is formulated as follows:
\begin{equation}
	\label{eqn1}
	\begin{split}
		\hat{\bm{h}}^{(v_{i})} &= \sum_{v_{t}\in{\mathcal{P_G}(v_{i})}}\bm{h}^{(v_{t})}, \\
		\bm{h}^{(v_{i})} &= g(U\bm{f}^{(v_{i})} + W\hat{\bm{h}}^{(v_{i})} + b),
	\end{split}
\end{equation}
where $\mathcal{P_G}(v_{i})$ is the direct predecessor set of $v_{i}$ in graph $\mathcal{G}$. $\hat{\bm{h}}^{(v_{i})}$ summarizes the information of the predecessors of $v_{i}$. The weight matrices $U$, $W$ are shared across predecessors in $\mathcal{P_G}(v_{i})$. $b$ is a learnable bias vector. $g(\cdot)$ is a nonlinear activation function (we use ReLU).

\subsubsection{Recurrent Feature Propagation Head}
In this section, we describe the pipeline of the RFP-Head for embedding context into local features (see Fig.~\ref{fig3} (d)). 
The RFP-Head receives input from the final layer of the shared encoder.
Its workflow mainly comprises the following steps (see the flowchart in Fig.~\ref{fig5}).

\begin{figure}[t]
	\centering
	\includegraphics[width=3.4in]{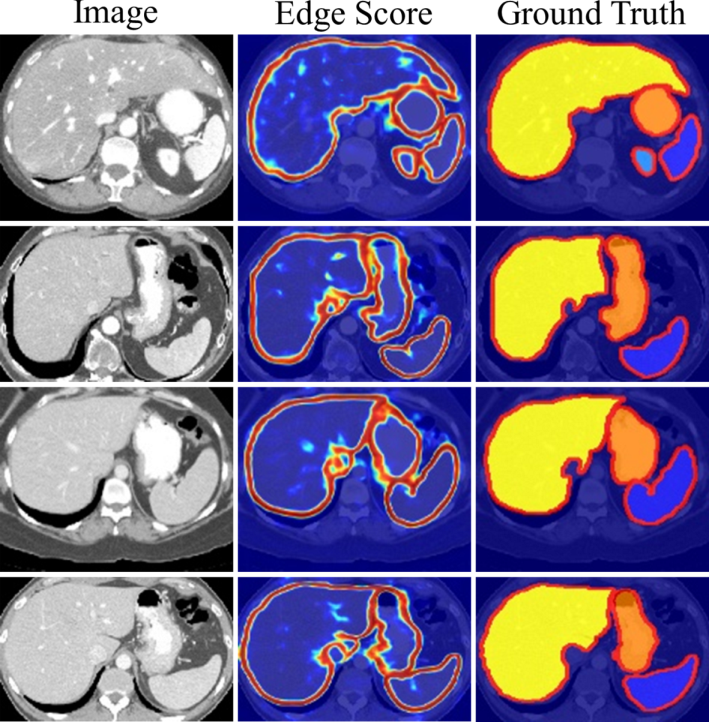}
	\caption{Qualitative illustration of the inferred edge likelihood maps generated by the edge detection sub-network. The red contours of organs in the third column are the edge ground truth.}
	\label{fig6}
\end{figure}

We denote the input features and corresponding hidden features as $\bm{E^{(5)}}, \bm{H}\in{\mathbb{R}^{C\times{H}\times{W}\times{D}}}$, where $C$ denotes the number of channels and $H\times{W}\times{D}$ denotes the spatial resolution.
Instead of directly using 3D-RNNs to process the whole 3D feature maps in cuboid order which is time-consuming (discussed below) and can create overlapping contexts~\cite{stollenga2015parallel}, we employ $D$ layers of UCG to independently perform context aggregation over $D$ feature maps. This design can be effective and amenable based on the insight that each of the $D$ feature maps from $\bm{E^{(5)}}$ has a large 3D receptive field thus capturing rich intra- and inter-slice contexts, critical for modeling contextual dependencies. In addition, it can be more efficient and suffers less from redundant contexts via the use of independent UCG-layers.
In our pratice, first, the feature $\bm{E^{(5)}}$ is split into $D$ feature maps. 
Second, these feature maps are iteratively input to the RFP-Module.
Third, the RFP-Module propagates the feature vectors $\bm{f}^{(v_i)}$ via DAGs with recurrent connections, to compute hidden feature vectors $\bm{h}^{(v_i)}$ (refer to the RFP-Module in Fig.~\ref{fig5} and Equation~(\ref{eqn1})). 
Fourth, the $D$ hidden maps generated by the RFP-Module are concatenated to reconstruct the hidden features $\bm{H}$. 
Note that we use independent weight matrices for each of the $D$ feature maps.
To facilitate gradient propagation, we introduce a residual connection producing the final enhanced features (see Fig.~\ref{fig3} (d)), which are passed through a $1\times{1}\times{1}$ convolution to generate the likelihood logits that capture wide-range context, beneficial for semantic segmentation.

In comparison with previous literature that used 3D-RNNs for volumetric medical image segmentation~\cite{stollenga2015parallel,andermatt2016multi}, our RFP-Head has multiple advantages.
First, our RFP-Module entails less computational costs by computing hidden features in an efficient manner.
Given dimension $d$, the $d$-dimension RNNs require $2^d$ hidden layers starting in every corner of the feature maps and scanning in opposite directions.
Thus 3D-RNNs generally compute eight hidden layers scanning through volumes, with a computational cost of 
$c^{\prime}\cdot{2^3}\cdot{(H\cdot{W}\cdot{D})}$.
In contrast, our RFP-Module iteratively processes the $D$ feature maps, thus with a computational cost of 
$c\cdot{2^2}\cdot{(H\cdot{W})}\cdot{D}$.
$c$ and $c^{\prime}$ denote computational costs of recurrent connections which grow linearly as cardinality of the direct predecessor set, thus $c < c^{\prime}$.
In such a way, our design requires at least two times less computational costs allowing the efficient processing of 3D feature maps.

Second, our RFP-Head can effectively capture rich contextual dependencies. 
As the $D$ feature maps are from the final layer of the encoder, each of them 
aggregates rich intra- and inter-slice semantic information.
By propagating and harvesting local features through DAGs, our RFP-Head can naturally model rich spatial dependencies over image regions, taking account of spatial arrangement of image units.
Furthermore, our RFP-Head receives direct back-propagation gradients from the output loss function, which enables effective parameter optimization and fast convergence.

\subsection{Edge Knowledge Integration}
\subsubsection{Edge Feature Extraction}
The task-specific edge detector exploits the intermediate features in encoder, to learn robust edge priors tuned for semantic segmentation. 
Considering the output quality and computation efficiency, we integrate $E^{(2)}$ and $E^{(5)}$ to provide effective edge representation.
The low-level $E^{(2)}$ is capable to provide local edge information, while the high-level feature $E^{(5)}$ has the largest receptive field thus is more spatial invariant, which can provide rough yet robust understanding of organ structures. 

To be specific, we first apply a convolutional unit to contract the channel number of the $E^{(5)}$ and upsample it by a factor of eight (see Fig.~\ref{fig3} (b)). 
Then, the adapted $E^{(5)}$ and the $E^{(2)}$ are added together. 
We use two consecutive convolutional units with $3\times{3}\times{3}$ kernel size to encode the edge representation $F_{edge}$.
A convolutional unit with kernel size of 1$\times$1$\times$1 and one output channel is applied to yield the edge likelihood maps. 
Some representative edge maps are illustrated in Fig.~\ref{fig6}.

The edge detection learning process is supervised by the reference edge maps $P$.
Specifically, we apply Canny detector~\cite{canny1986computational} on segmentation ground-truth to delineate edges of different organs (see the third column in Fig.~\ref{fig6}).
In the training process, there exists severe class imbalance problem between the background and organ edges. 
Namely, the background class occupies large percentage of voxels, while the edge class is relatively rare.
In order to mitigate this problem, we introduce a weighted binary cross-entropy loss $\mathcal{L}_{edge}$: 
\begin{equation}
	\begin{split}
		\mathcal{L}_{edge} = -\frac{1}{N} [ &\sum_{i\in{P_+}} \alpha \log{(\tilde{p}_i)} \\
		+ &\sum_{i\in{P_-}} (1-\alpha) \log{(1-\tilde{p}_i)} ], 
	\end{split}
\end{equation}
where $N$ is the total number of voxels in the $P$; 
$P_+$ and $P_-$ denote the edge and background voxel set, respectively; 
$\tilde{p}_{i}$ is the edge probability at voxel $i$; 
$\alpha$ is a weight scalar.
We denote the number of edge voxels as $|P_+|$ and background voxels as $|P_-|$.
The weight scalar $\alpha$ is calculated by $\frac{|P_-|}{|P_+| + |P_-|}$ and assigned to the loss term for edge class.
By doing so, the edge class reasonably has a larger weight than the background class, thus gains more attention, resulting in more distinct edge representation for semantic segmentation.

\subsubsection{Edge Skip-Connections}

\begin{figure}[t]
	\centering
	\includegraphics[width=3.5in]{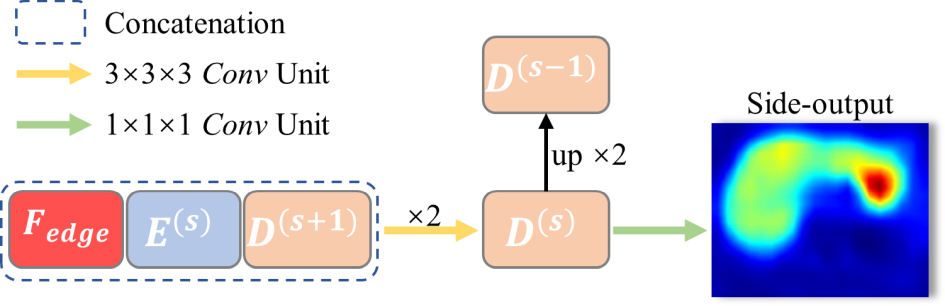}
	\caption{Decoder blocks with edge skip-connections (ESCs). The edge feature $F_{edge}$ is propagated to each decoder block. After merging the edge feature, the decoder feature $D^{(s)}$ is processed by a $1\times{1}\times{1}$ convolutional unit to generate the segmentation likelihood logits, i.e., side-output.}
	\label{fig7}
\end{figure}

After obtaining the $F_{edge}$, we aim to use the edge knowledge to guide the segmentation predictions.
One straightforward way is to combine the $F_{edge}$ with the final decoder feature $D^{(1)}$.
However, we argue that better performance can be obtained by fusing the edge feature with multi-level decoder features.
Here, we introduce the edge skip-connections (ESCs) to propagate the $F_{edge}$ to multi-level decoder blocks, enabling the edge feature streams to work at different resolutions.
Specifically, at each decoder block, previous feature $D^{(s+1)}$ is merged with the edge feature $F_{edge}$ propagated by ESCs, and the same-scale encoder feature $E^{(s)}$, yielding the decoder feature maps $D^{(s)}$ (see Fig.~\ref{fig7}). $s$ indicates the stage of decoder block.
Such ESCs explicitly aggregates edge priors to multi-level decoder features, which is beneficial to learn a hierarchy of features explicitly modeling complementarity between organs' interiors and edges.

To improve discrimination capability of multi-level features, deep supervision is further added at each decoder blocks.
Specifically, each decoder feature is processed by a convolution with $1\times{1}\times{1}$ kernel size and $K$ output channels generating the intermediate segmentation likelihood logits, namely side-output. 
$K$ is the number of classes.
All side-outputs are concatenated together to jointly determine the final segmentation predictions (see Fig.~\ref{fig3} (c)). 

\section{Experimental Methods}
\subsection{Datasets}
We use 90 CT scans from two public multi-organ datasets: 43 scans from the Pancreas-CT dataset~\cite{roth2015deeporgan} and 47 scans from the Beyond the Cranial Vault (BTCV) segmentation challenge\footnote{https://www.synapse.org/\#!Synapse:syn3193805}. 
Both of the datasets consist of eight manually labeled organs provided by~\cite{gibson2018automatic}, including: spleen (Spl.), left kidney (L. Kid.), gallbladder (Gallb.), esophagus (Esoph.), liver (Liv.), stomach (Stom.), pancreas (Panc.) and duodenum (Duod.).
The CT scans have transversal resolutions of $512\times{512}$ pixels with pixel sizes from $0.6\sim0.9~mm$ and slice thickness between $0.5\sim5.0~mm$.

\subsection{Evaluation Metrics}
The metrics employed to quantitatively evaluate the segmentation accuracy included 
Dice similarity coefficient (DSC),
average symmetric surface distance (ASSD),
and symmetric 95\% Hausdorff distance (95HD)~\cite{wang2019deep}.
DSC measures the relative volumetric overlap between the predicted and reference segmentations.
ASSD determines the average distance between the surfaces of the predicted and reference segmentations in 3D.
95HD is similar to ASSD but more sensitive to the localized disagreement as it determines the 95\textit{th} percentile of all calculated Hausdorff distances.
Four-fold cross-validation was conducted to evaluate the segmentation performance.
A better segmentation shall have smaller ASSD and 95HD, and larger value of DSC metric.



\subsection{Implementation Details}
All CT volumes were resampled to the size of $160\times{160}\times{64}$. 
The intensity values were clipped to an interval of $[-250, 200]$ HU, and then normalized to zero mean and unit variance.
During training, random rotation and random flipping were applied for data augmentation. 

The segmentation model was implemented on the open source platform Pytorch.
Gradient updates were computed using the mini-batch size of two samples.
The model was trained using batch normalization, deep supervision~\cite{lee2015deeply} and the Adam optimizer with weight decay of $3e-4$.
The base learning rate $lr_b$ was set as $1e-3$.
The total number of training epochs was 400.
We used a ``poly" learning rate decay strategy, where the current learning rate $lr_c$ depended on the $lr_b$ and a scale factor related to the training epochs: $lr_c=lr_b\times{(1-\frac{epoch}{total\_epoch})^{0.9}}$.
All experiments were conducted on a NVIDIA RTX 2080 Ti GPU with 11 GB memory.

\subsection{Comparison with Existing Models}
We compared the segmentation performance of the proposed network to those of seven deep-learning-based models, including two baseline models~\cite{cciccek20163d, milletari2016v}, two well-established models for medical image segmentation~\cite{oktay2018attention, zhou2019unet++}, and three state-of-the-art abdominal organ segmentation models~\cite{zhang2020block, chen2021transunet, gibson2018automatic}.

\subsubsection{Baseline Models}

\begin{itemize}
	\item UNet~\cite{cciccek20163d}: is an encoder-decoder architecture with skip connections.
	\item VNet~\cite{milletari2016v}: can be seen as a variant of the UNet architecture with block-wise residual connections~\cite{he2016deep}.
\end{itemize}

\subsubsection{Well-established Medical Image Segmentation Models}

\begin{itemize}
	\item AttentionUNet~\cite{oktay2018attention}: integrates attention gates into the UNet model, which filter features propagated through the skip-connections\footnote{We implemented the AttentionUNet using the publicly available implementation (https://github.com/ozan-oktay/Attention-Gated-Networks).}.
	
	\item UNet++~\cite{zhou2019unet++}: is an ensemble of UNets with varying depths, which share the same encoder but have their own decoders and consists of densely connected skip-connections\footnote{We implemented the UNet++ based on the publicly available implementation (https://github.com/MrGiovanni/UNetPlusPlus).}.
\end{itemize}

\subsubsection{State-of-the-art Organ Segmentation Models}

\begin{itemize}
	\item CascadedVNet~\cite{zhang2020block}: is a cascaded network, which introduces extra block-wise skip-connections between two cascaded VNets and uses stacked and inception-like convolutions at encoder blocks\footnote{We re-implemented the CascadedVNet based on the VNet backbone (https://github.com/faustomilletari/VNet). Supervision signals were imposed on both of the first and second stage networks.}.
	
	\item TransUNet~\cite{chen2021transunet}: is a recent work that combines Transformer~\cite{vaswani2017attention} and UNet~\cite{ronneberger2015u} for 2D medical image segmentation. 
	Specifically, the TransUNet introduces global contexts into the encoder via the usage of 12 Vision Transformers (ViT) layers~\cite{dosovitskiy2020image}\footnote{We implemented the TransUnet using publicly available repository (https://github.com/Beckschen/TransUNet). Following the default setting in TransUNet~\cite{chen2021transunet}, we used the patch size of $16\times{16}$ pixels. We set the input resolution as $160\times160$ pixels. The mini-batch size was set as 24 samples. In inference phase, all 3D volumes were predicted in a slice-by-slice fashion and reconstructed by stacking the predicted 2D slices.}.
	
	\item DenseVNet~\cite{gibson2018automatic}: is based on the VNet architecture but additionally incorporates batch-wise spatial dropout, dense feature stacks, dilated convolutions and a spatial prior for abdominal organ segmentation\footnote{We implemented the DenseVNet based on the publicly available implementation (https://github.com/NifTK/NiftyNet).}.
\end{itemize}

We sufficiently implemented and trained all comparing models.

\begin{table*}[t]
	\centering
	\setlength{\extrarowheight}{1pt}
	\caption{Quantitative comparison between our method and other cutting-edge methods on the abdominal multi-organ segmentation dataset. \textbf{Values in boldface denote the best results.}}
	\label{model_comparison}
	\begin{tabular}{c|llllllll|c}
		\hline
		\hline
		Methods & ~~Spl. & ~~L. Kid. & ~~Gallb. & ~~Esoph. & ~~Liv. & ~~Stom. & ~~Panc. & ~~Duod. & All \\ \hline
		\multicolumn{1}{c}{} & \multicolumn{8}{c}{Dice coefficient (mean$\pm$sd., \%)} &                                                                           \\ \hline
		UNet~\cite{cciccek20163d} & 91.9{\tiny $\pm$15.3}$^*$ & 91.6{\tiny $\pm$15.8} & 71.1{\tiny $\pm$28.6}$^*$ & 72.4{\tiny $\pm$13.0}$^*$ & 95.4{\tiny $\pm$4.8}$^*$ & 86.5{\tiny $\pm$12.6}$^*$ & 74.6{\tiny $\pm$14.0}$^*$ & 65.1{\tiny $\pm$15.6}$^*$ & 81.1{\tiny $\pm$10.8}$^*$\\
		VNet~\cite{milletari2016v} & 91.5{\tiny $\pm$13.9}$^*$ & 88.3{\tiny $\pm$21.4}$^*$ & 70.8{\tiny $\pm$27.7}$^*$ & 71.9{\tiny $\pm$12.1}$^*$ & 94.6{\tiny $\pm$5.6}$^*$ & 84.8{\tiny $\pm$11.7}$^*$ & 73.5{\tiny $\pm$11.8}$^*$ & 62.8{\tiny $\pm$13.8}$^*$ & 79.8{\tiny $\pm$10.7}$^*$ \\ \hline
		AttentionUNet~\cite{oktay2018attention} & 93.3{\tiny $\pm$11.1}$^*$ & 91.3{\tiny $\pm$16.5} & 73.3{\tiny $\pm$26.2}$^*$ & 72.8{\tiny $\pm$12.9}$^*$ & 95.5{\tiny $\pm$4.7}$^*$ & 88.0{\tiny $\pm$8.3}$^*$ & 76.2{\tiny $\pm$13.0}$^*$ & 66.3{\tiny $\pm$12.8} & 82.1{\tiny $\pm$10.4}$^*$ \\
		UNet++~\cite{zhou2019unet++} & 92.9{\tiny $\pm$11.3}$^*$ & 91.6{\tiny $\pm$15.6} & 73.3{\tiny $\pm$27.4}$^*$ & 74.1{\tiny $\pm$12.1}$^*$ & 95.5{\tiny $\pm$4.5} & 87.0{\tiny $\pm$11.8}$^*$ & 76.6{\tiny $\pm$13.0}$^*$ & 67.7{\tiny $\pm$13.3} & 82.3{\tiny $\pm$9.9}$^*$ \\
		TransUNet~\cite{chen2021transunet} &\textbf{94.2{\tiny$\pm$ 8.1}} &92.3{\tiny$\pm$ 12.5} &72.1{\tiny$\pm$ 28.1}             & 72.5{\tiny$\pm$ 13.8}$^*$ &95.4{\tiny$\pm$ 5.3} &88.1{\tiny$\pm$ 8.3}$^*$ &75.7{\tiny$\pm$ 10.7}$^*$ &64.6{\tiny$\pm$ 14.9}$^*$ &81.9{\tiny $\pm$11.2}$^*$      \\
		CascadedVNet~\cite{zhang2020block} & 92.6{\tiny $\pm$13.1}$^*$ & 89.7{\tiny $\pm$19.6}$^*$ & 74.3{\tiny $\pm$26.5}$^*$ & 73.5{\tiny $\pm$11.6} & 95.1{\tiny $\pm$5.3}$^*$ & 85.7{\tiny $\pm$12.8}$^*$ & 75.3{\tiny $\pm$11.2}$^*$ & 64.6{\tiny $\pm$13.3}$^*$ & 81.4{\tiny $\pm$10.2}$^*$ \\
		DenseVNet~\cite{gibson2018automatic} & 93.4{\tiny $\pm$10.5}$^*$ & 90.3{\tiny $\pm$17.8} & 73.4{\tiny $\pm$27.1}$^*$ & 72.8{\tiny $\pm$12.5}$^*$ & 95.5{\tiny $\pm$4.5}$^*$ & 87.7{\tiny $\pm$8.5}$^*$ & 75.6{\tiny $\pm$11.7}$^*$ & 64.6{\tiny $\pm$14.8}$^*$ & 81.6{\tiny $\pm$10.7}$^*$ \\
		Proposed & \textbf{94.2{\tiny $\pm$9.4}} & \textbf{92.4{\tiny $\pm$16.9}} & \textbf{74.5{\tiny $\pm$27.4}} & \textbf{74.8{\tiny $\pm$11.7}} & \textbf{95.7{\tiny $\pm$4.4}} & \textbf{89.3{\tiny $\pm$8.1}} & \textbf{78.1{\tiny $\pm$11.7}} & \textbf{67.9{\tiny $\pm$14.0}} & \textbf{83.4{\tiny $\pm$10.0}}~ \\ \hline
		
		\multicolumn{1}{c}{} & \multicolumn{8}{c}{Average symmetric surface distance (mean$\pm$sd., $mm$)} &                                                         \\ \hline
		UNet~\cite{cciccek20163d} & 1.69{\tiny $\pm$6.87}$^*$ & 0.38{\tiny $\pm$0.50} & 3.25{\tiny $\pm$11.08}$^*$ & 0.83{\tiny $\pm$0.81}$^*$ & 0.46{\tiny $\pm$2.36}$^*$ & 1.03{\tiny $\pm$1.68}$^*$ & 1.24{\tiny $\pm$2.03}$^*$ & 2.71{\tiny $\pm$3.03} & 1.45{\tiny $\pm$0.98}$^*$ \\
		VNet~\cite{milletari2016v} & 2.83{\tiny $\pm$13.13}$^*$ & 1.02{\tiny $\pm$3.20}$^*$ & 4.20{\tiny $\pm$12.48}$^*$ & 0.75{\tiny $\pm$0.56}$^*$ & 0.55{\tiny $\pm$2.37}$^*$ & 1.21{\tiny $\pm$1.80}$^*$ & 1.11{\tiny $\pm$1.37}$^*$ & 2.72{\tiny $\pm$3.16}$^*$ & 1.80{\tiny $\pm$1.21}$^*$ \\ \hline
		AttentionUNet~\cite{oktay2018attention} & 1.16{\tiny $\pm$9.79} & \textbf{0.18{\tiny $\pm$0.70}} & 3.41{\tiny $\pm$8.58} & 0.72{\tiny $\pm$0.55} & 0.48{\tiny $\pm$2.07} & 0.66{\tiny $\pm$1.46}$^*$ & 1.16{\tiny $\pm$2.25}$^*$ & 2.88{\tiny $\pm$2.80} & 1.30{\tiny $\pm$1.12}$^*$ \\
		UNet++~\cite{zhou2019unet++} & 1.32{\tiny $\pm$5.24}$^*$ & 0.19{\tiny $\pm$0.61} & 3.69{\tiny $\pm$12.56}$^*$ & 0.70{\tiny $\pm$0.62} & 0.44{\tiny $\pm$2.28} & 0.92{\tiny $\pm$1.50}$^*$ & 1.08{\tiny $\pm$1.94}$^*$ & 2.57{\tiny $\pm$2.60} & 1.36{\tiny $\pm$1.11}$^*$ \\
		TransUNet~\cite{chen2021transunet} & 2.70{\tiny $\pm$7.99}$^*$ & 2.84{\tiny $\pm$6.31}$^*$ & 4.39{\tiny $\pm$14.71}$^*$ & 1.89{\tiny $\pm$1.32}$^*$ & 2.31{\tiny $\pm$4.23}$^*$ & 3.09{\tiny $\pm$2.47}$^*$ & 2.41{\tiny $\pm$1.78}$^*$ & 4.61{\tiny $\pm$3.38}$^*$ & 3.03{\tiny $\pm$0.92}$^*$ \\
		CascadedVNet~\cite{zhang2020block} & 0.80{\tiny $\pm$3.62} & 0.40{\tiny $\pm$1.11}$^*$ & 4.15{\tiny $\pm$13.81}$^*$ & 0.71{\tiny $\pm$0.58} & 0.48{\tiny $\pm$2.34}$^*$ & 1.26{\tiny $\pm$2.40}$^*$ & 1.25{\tiny $\pm$2.00}$^*$ & 2.58{\tiny $\pm$2.36} & 1.45{\tiny $\pm$1.21}$^*$ \\
		DenseVNet~\cite{gibson2018automatic} & 1.00{\tiny $\pm$3.89}$^*$ & 1.03{\tiny $\pm$4.84} & 2.26{\tiny $\pm$6.82}$^*$ & 0.82{\tiny $\pm$1.27} & 0.44{\tiny $\pm$2.23} & 1.02{\tiny $\pm$2.08}$^*$ & 1.27{\tiny $\pm$2.36} & 3.73{\tiny $\pm$8.16}$^*$ & 1.45{\tiny $\pm$0.99}$^*$ \\
		Proposed & \textbf{0.69{\tiny $\pm$3.93}} & 0.38{\tiny $\pm$1.87} & \textbf{1.99{\tiny $\pm$7.92}} & \textbf{0.62{\tiny $\pm$0.56}} & \textbf{0.40{\tiny $\pm$2.21}} & \textbf{0.64{\tiny $\pm$1.04}} & \textbf{1.07{\tiny $\pm$2.08}} & \textbf{2.45{\tiny $\pm$2.99}} & \textbf{1.03{\tiny $\pm$0.72}}~ \\ \hline
		
		\multicolumn{1}{c}{} & \multicolumn{8}{c}{95\% Hausdorff distance (mean$\pm$sd., $mm$)} &                                                          \\ \hline
		UNet~\cite{cciccek20163d} & 10.18{\tiny $\pm$37.34}$^*$ & 1.49{\tiny $\pm$5.26} & 10.72{\tiny $\pm$28.02} & 4.10{\tiny $\pm$3.40}$^*$ & 2.51{\tiny $\pm$15.68} & 6.36{\tiny $\pm$10.39} & 7.78{\tiny $\pm$13.28}$^*$ & 15.72{\tiny $\pm$15.68} & 7.36{\tiny $\pm$4.46}$^*$ \\
		VNet~\cite{milletari2016v} & 13.89{\tiny $\pm$55.01}$^*$ & 5.94{\tiny $\pm$20.87}$^*$ & 15.97{\tiny $\pm$41.78} & 4.08{\tiny $\pm$3.06}$^*$ & 3.46{\tiny $\pm$16.07}$^*$ & 7.37{\tiny $\pm$10.59}$^*$ & 6.41{\tiny $\pm$7.84}$^*$ & 15.30{\tiny $\pm$15.90} & 9.05{\tiny $\pm$4.82}$^*$ \\ \hline
		AttentionUNet~\cite{oktay2018attention} & 9.79{\tiny $\pm$43.67}$^*$ & \textbf{1.02{\tiny $\pm$3.02}} & 13.30{\tiny $\pm$35.04} & 3.78{\tiny $\pm$2.58}$^*$ & 2.60{\tiny $\pm$15.21} & 5.84{\tiny $\pm$11.96} & 6.40{\tiny $\pm$10.88}$^*$ & 14.79{\tiny $\pm$14.76} & 7.19{\tiny $\pm$4.68}$^*$ \\
		UNet++~\cite{zhou2019unet++} & 7.22{\tiny $\pm$29.13}$^*$ & 1.22{\tiny $\pm$5.60} & 11.27{\tiny $\pm$31.74} & 3.93{\tiny $\pm$3.54} & 2.53{\tiny $\pm$15.63} & 5.43{\tiny $\pm$9.02} & 6.68{\tiny $\pm$12.55}$^*$ & 15.39{\tiny $\pm$14.32} & 6.71{\tiny $\pm$4.38}$^*$ \\
		TransUNet~\cite{chen2021transunet} & 13.35{\tiny $\pm$35.52}$^*$ & 15.91{\tiny $\pm$34.60}$^*$ & 11.70{\tiny $\pm$23.89}$^*$ & 7.08{\tiny $\pm$4.80}$^*$ & 11.81{\tiny $\pm$31.16}$^*$ & 16.39{\tiny $\pm$17.97}$^*$ & 9.43{\tiny $\pm$7.33}$^*$ & 20.32{\tiny $\pm$14.74}$^*$ & 13.25{\tiny $\pm$3.94}$^*$ \\
		CascadedVNet~\cite{zhang2020block} & 5.04{\tiny $\pm$25.85}$^*$ & 2.82{\tiny $\pm$12.46} & 13.08{\tiny $\pm$35.83} & 3.93{\tiny $\pm$3.56}$^*$ & 2.69{\tiny $\pm$15.48} & 8.67{\tiny $\pm$19.80}$^*$ & 8.13{\tiny $\pm$14.14}$^*$ & 15.45{\tiny $\pm$13.98} & 7.48{\tiny $\pm$4.47}$^*$ \\
		DenseVNet~\cite{gibson2018automatic} & 6.51{\tiny $\pm$27.02} & 5.61{\tiny $\pm$24.26} & 10.12{\tiny $\pm$24.79} & 4.62{\tiny $\pm$7.89}$^*$ & \textbf{2.46{\tiny $\pm$15.50}} & 7.96{\tiny $\pm$20.49} & 7.18{\tiny $\pm$13.43}$^*$ & 20.31{\tiny $\pm$31.44}$^*$ & 8.10{\tiny $\pm$5.08}$^*$ \\
		Proposed & \textbf{3.56{\tiny $\pm$20.51}} & 2.59{\tiny $\pm$12.03} & \textbf{6.66{\tiny $\pm$18.23}} & \textbf{3.58{\tiny $\pm$3.48}} & \textbf{2.46{\tiny $\pm$15.75}} & \textbf{4.55{\tiny $\pm$11.10}} & \textbf{5.51{\tiny $\pm$13.52}} & \textbf{14.26{\tiny $\pm$15.31}} & \textbf{5.40{\tiny $\pm$3.61}}~ \\ \hline
		\hline
	\end{tabular}
	
	\vspace{1mm}
	{\raggedright \textit{~~~$^*$ indicates the results are statistically different with ours ($p$-value$<$0.05).} \par}
\end{table*}

\subsection{Analytical Ablation Study}
We further conducted ablation analyses to investigate the efficacy of our architecture designs.

\subsubsection{Effect of Devised Components}
To isolate and quantify the contribution of each component of the proposed architecture, we conducted a series of ablation experiments, wherein we altered key designs of the architecture: edge detection sub-network (ED), edge skip-connections (ESCs) and recurrent feature propagation head (RFP-Head).

\subsubsection{Different Number of DAGs}
We compared variants of our RFP-Head, each of which used different number of DAGs (i.e., 0, 1, 2, 4 DAGs) to approximate the UCG.
Note that the segmentation network didn't have context aggregation module in the case where no DAG was used.

\subsubsection{Different Number of Edge Skip-connections}
To investigate the effective design of the ESCs, we varied the number of ESCs to be 0, 1, 2 or 4.
Note that in the one ESC setting, the edge features were propagated to the final decoder block; where in the two ESCs setting, the edge features were propagated to the last two decoder blocks.

\subsubsection{Neighborhood Connectivity of UCG}
\begin{figure}[t]
	\centering
	\includegraphics[width=3.4in]{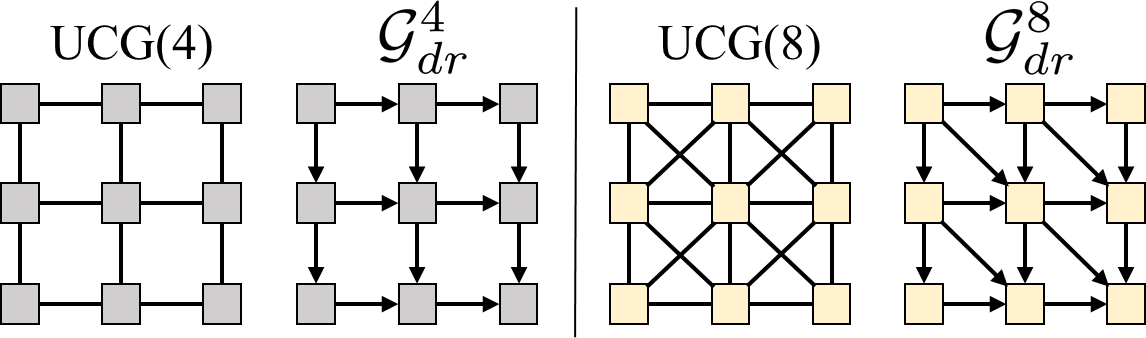}
	\caption{UCGs with 4 and 8 neighborhood system and their induced DAGs scanning down and right. Comparing to the $\mathcal{G}_{dr}^{4}$, the $\mathcal{G}_{dr}^{8}$ incorporates additional diagonal connections.}
	\label{fig9}
\end{figure}

To explore the impact of neighborhood connectivity, we conducted experiments using RFP-Head with UCG of four and eight neighborhood system (refer to Fig.~\ref{fig9}).
Comparing to the $\mathcal{G}_{dr}^{4}$, the $\mathcal{G}_{dr}^{8}$ incorporated additional diagonal connections.

%
%

\begin{figure*}[]
	\centering
	\includegraphics[width=6.5in]{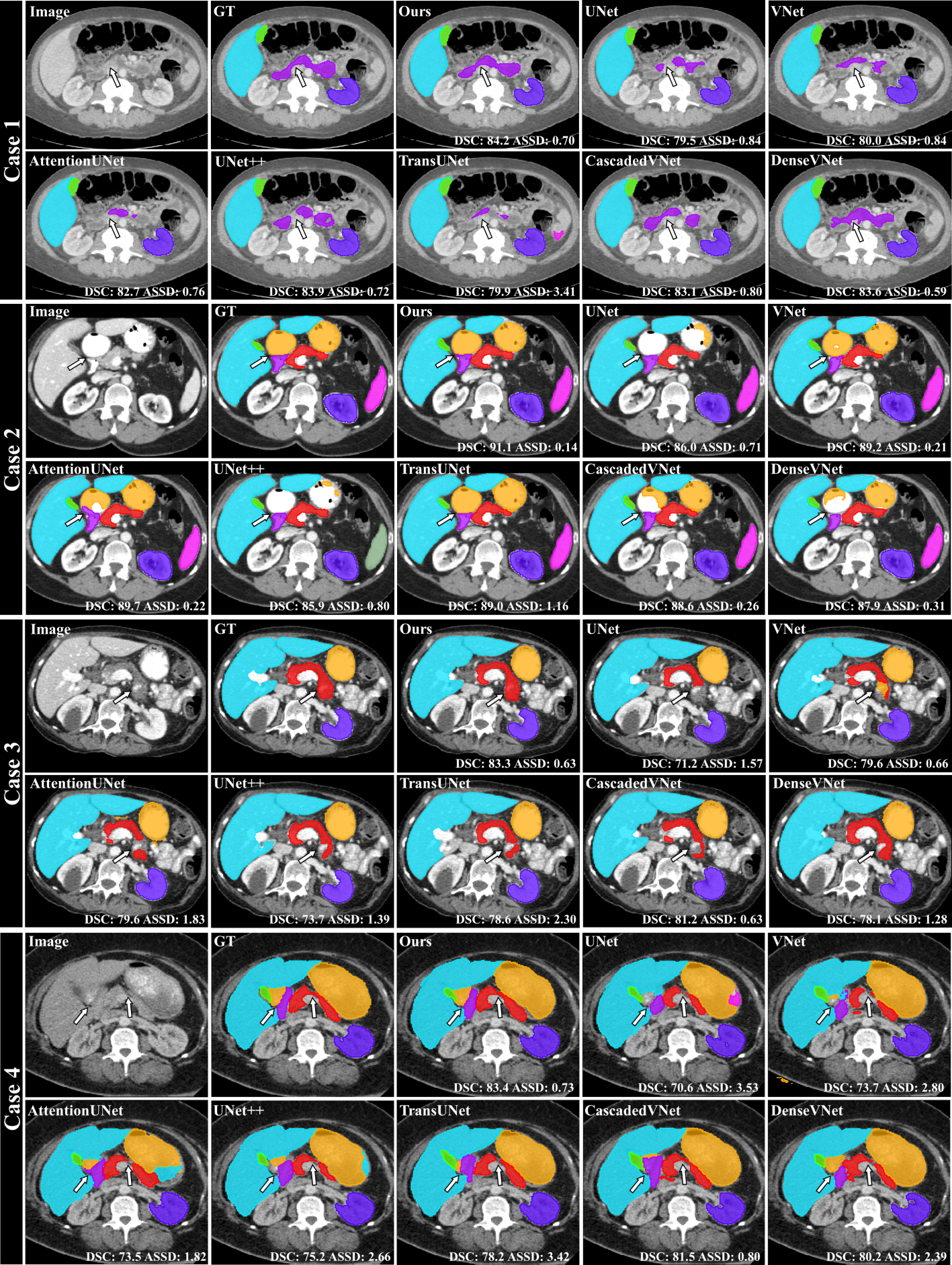}
	\caption{Qualitative illustration of predicted segmentations in axial view. The overall average DSC ($\%$) and ASSD ($mm$) of predicted volumetric segmentations are provided at the bottom of each sample. The white arrows indicate areas that our method achieved satisfactory performance. The following organs are displayed above: pancreas (red), duodenum (purple), stomach (orange), liver (cyan), left-kidney (blue), gallbladder (green) and spleen (magenta).}
	\label{fig8}
\end{figure*}

\section{Results and Discussion}

\subsection{Comparison with Existing Models}
Table~\ref{model_comparison} reports the numerical results of all methods.
The average Dice scores across all organs of the UNet and VNet baselines are 81.1$\%$ and 79.8$\%$, respectively.
Note that we used the same abdominal CT dataset as the DenseVNet~\cite{gibson2018automatic}. 
The average Dice score reported in~\cite{gibson2018automatic} is 81.5$\%$, while our re-implementation result is 81.6$\%$.

\begin{table*}[t]
	\centering
	\setlength{\extrarowheight}{1pt}
	\caption{Ablation analyses of the proposed network architecture. \textbf{Values in boldface denote the best results.}}
	\label{ablation_study}
	\begin{tabular}{>{\centering}m{1cm}|>{\centering}m{0.5cm}|>{\centering}m{0.7cm}|>{\centering}m{0.5cm}|cccccccc|c}
		\hline
		\hline
		\multicolumn{4}{c|}{Modules}   & Spl. & L. Kid. & Gallb. & Esoph. & Liv. & Stom. & Panc. & Duod.  & All \\ \hline
		Backbone     & ED           & ESCs          & RFP     & \multicolumn{9}{c}{Dice Coefficient (mean$\pm$sd., \%)}                                                     \\ \hline
		$\checkmark$ &              &              &              &91.9{\tiny$\pm$15.3} &91.6{\tiny$\pm$15.8} &71.1{\tiny$\pm$28.6} &72.4{\tiny$\pm$13.0} &95.4{\tiny$\pm$4.8} &86.5{\tiny$\pm$12.6} &74.6{\tiny$\pm$14.0} &65.1{\tiny$\pm$15.6} & 81.1{\tiny $\pm$10.8} \\
		$\checkmark$ & $\checkmark$ &              &              &92.8{\tiny$\pm$12.6} &91.2{\tiny$\pm$15.7} &72.2{\tiny$\pm$27.9} &72.3{\tiny$\pm$12.7} &95.3{\tiny$\pm$5.2} &88.0{\tiny$\pm$9.5}  &75.7{\tiny$\pm$13.6} &65.1{\tiny$\pm$15.0} & 81.6{\tiny $\pm$10.8}\\
		$\checkmark$ & $\checkmark$ & $\checkmark$ &              &93.9{\tiny$\pm$10.4} &91.5{\tiny$\pm$19.3} &72.6{\tiny$\pm$28.3} &73.5{\tiny$\pm$12.1} &95.6{\tiny$\pm$4.5} &88.3{\tiny$\pm$9.7} &77.1{\tiny$\pm$12.6} &65.5{\tiny$\pm$15.7} & 82.3{\tiny $\pm$10.7} \\
		$\checkmark$ &              &              & $\checkmark$ &94.0{\tiny$\pm$10.6} &91.7{\tiny$\pm$19.4} &73.2{\tiny$\pm$27.1} &73.4{\tiny$\pm$13.5} &\textbf{95.7{\tiny$\pm$4.4}} &89.1{\tiny$\pm$8.0} &76.0{\tiny$\pm$12.7} &67.0{\tiny$\pm$15.1} & 82.5{\tiny $\pm$10.5} \\
		$\checkmark$ & $\checkmark$ & $\checkmark$ & $\checkmark$ &\textbf{94.2{\tiny$\pm$9.4}}  &\textbf{92.4{\tiny$\pm$16.9}} &\textbf{74.5{\tiny$\pm$27.4}} &\textbf{74.8{\tiny$\pm$11.7}} &\textbf{95.7{\tiny$\pm$4.4}} &\textbf{89.3{\tiny$\pm$8.1}} &\textbf{78.1{\tiny$\pm$11.7}} &\textbf{67.9{\tiny$\pm$14.0}} & \textbf{83.4{\tiny $\pm$10.0}} \\ \hline
		Backbone     & ED           & ESCs          & RFP     & \multicolumn{9}{c}{Average symmetric surface distance (mean$\pm$sd., $mm$)} \\ \hline
		$\checkmark$ &              &              &              &1.69{\tiny$\pm$6.87} &0.38{\tiny$\pm$0.50} &3.25{\tiny$\pm$11.08} &0.83{\tiny$\pm$0.81} &0.46{\tiny$\pm$2.36} &1.03{\tiny$\pm$1.68} &1.24{\tiny$\pm$2.03} &2.71{\tiny$\pm$3.03} & 1.45{\tiny $\pm$0.98} \\
		$\checkmark$ & $\checkmark$ &              &              &2.23{\tiny$\pm$13.59} &0.53{\tiny$\pm$2.61} &2.81{\tiny$\pm$8.46} &0.76{\tiny$\pm$0.61} &0.45{\tiny$\pm$2.22} &0.90{\tiny$\pm$1.54} &1.13{\tiny$\pm$2.22} &2.99{\tiny$\pm$3.24} & 1.47{\tiny $\pm$0.97} \\
		$\checkmark$ & $\checkmark$ & $\checkmark$ &              &1.07{\tiny$\pm$5.42} &0.49{\tiny$\pm$1.73} &3.87{\tiny$\pm$16.71} &0.83{\tiny$\pm$1.53} &0.37{\tiny$\pm$1.99} &0.78{\tiny$\pm$1.19} &\textbf{0.98{\tiny$\pm$2.42}} &3.09{\tiny$\pm$3.65} & 1.43{\tiny $\pm$1.21} \\
		$\checkmark$ &              &              & $\checkmark$ &1.50{\tiny$\pm$11.08} &\textbf{0.28{\tiny$\pm$1.30}} &3.15{\tiny$\pm$15.11} &0.69{\tiny$\pm$0.54} &\textbf{0.32{\tiny$\pm$1.93}} &\textbf{0.62{\tiny$\pm$0.99}} &1.05{\tiny$\pm$1.73} &2.57{\tiny$\pm$3.22} & 1.27{\tiny $\pm$1.00} \\
		$\checkmark$ & $\checkmark$ & $\checkmark$ & $\checkmark$ &\textbf{0.69{\tiny$\pm$3.93}} &0.38{\tiny$\pm$1.87} &\textbf{1.99{\tiny$\pm$7.92}} &\textbf{0.62{\tiny$\pm$0.56}} &0.40{\tiny$\pm$2.21} &0.64{\tiny$\pm$1.04} &1.07{\tiny$\pm$2.08} & \textbf{2.45{\tiny$\pm$2.99}} & \textbf{1.03{\tiny $\pm$0.72}} \\ \hline
		Backbone     & ED           & ESCs          & RFP     & \multicolumn{9}{c}{95\% Hausdorff distance (mean$\pm$sd., $mm$)}  \\ \hline
		$\checkmark$ &              &              &              &10.18{\tiny$\pm$37.34} &1.49{\tiny$\pm$5.26} &10.72{\tiny$\pm$28.02} &4.10{\tiny$\pm$3.40} &2.51{\tiny$\pm$15.68} &6.36{\tiny$\pm$10.39} &7.78{\tiny$\pm$13.28} &15.72{\tiny$\pm$15.68} &7.36{\tiny $\pm$4.46} \\
		$\checkmark$ & $\checkmark$ &              &              &9.51{\tiny$\pm$50.02} &2.76{\tiny$\pm$13.11} &13.25{\tiny$\pm$31.38} &4.10{\tiny$\pm$3.04} &2.74{\tiny$\pm$15.67} &5.57{\tiny$\pm$10.43} &6.87{\tiny$\pm$12.49} &17.09{\tiny$\pm$16.63} &7.75{\tiny $\pm$4.88} \\
		$\checkmark$ & $\checkmark$ & $\checkmark$ &              &6.27{\tiny$\pm$33.16} &3.05{\tiny$\pm$11.99} &10.26{\tiny$\pm$28.51} &4.74{\tiny$\pm$10.35} &2.35{\tiny$\pm$14.83} &4.95{\tiny$\pm$9.69} &\textbf{5.45{\tiny$\pm$11.55}} &18.34{\tiny$\pm$21.92} &6.93{\tiny $\pm$4.85} \\
		$\checkmark$ &              &              & $\checkmark$ &7.07{\tiny$\pm$50.24} &\textbf{1.46{\tiny$\pm$6.69}} &12.84{\tiny$\pm$40.39} &3.89{\tiny$\pm$3.42} &\textbf{1.46{\tiny$\pm$12.37}} &\textbf{3.87{\tiny$\pm$8.31}} &6.78{\tiny$\pm$12.77} &14.29{\tiny$\pm$17.27} &6.46{\tiny $\pm$4.55} \\
		$\checkmark$ & $\checkmark$ & $\checkmark$ & $\checkmark$ &\textbf{3.56{\tiny$\pm$20.51}} &2.59{\tiny$\pm$12.03} &\textbf{6.66{\tiny$\pm$18.23}} &\textbf{3.58{\tiny$\pm$3.48}} &2.46{\tiny$\pm$15.75} &4.55{\tiny$\pm$11.10} &5.51{\tiny$\pm$13.52} &\textbf{14.26{\tiny$\pm$15.31}} & \textbf{5.40{\tiny $\pm$3.61}} \\ \hline
		\hline
	\end{tabular}
	
	\vspace{1mm}
	{\raggedright \textit{~~~Backbone is the 3D UNet. ED denotes edge detection sub-network; ESCs denotes edge skip-connections; \\~~~RFP denotes recurrent feature propagation head.} \par}
\end{table*}

As shown in Table~\ref{model_comparison}, the proposed network consistently outperformed other comparing models on almost all the metrics.
Specifically, our network yielded higher Dice values for each organ than those of state-of-the-art models, and excluding the left kidney (discussed below), the proposed network yielded lower ASSD and 95HD values than others'.
Our network achieved an overall average Dice value of 83.4$\%$, exceeding UNet by 2.3$\%$, VNet by 3.6$\%$, AttentionUNet by 1.3$\%$, UNet++ by 1.1$\%$, TransUNet by 1.5$\%$, CascadedVNet by 2.0$\%$, and DenseVNet by 1.8$\%$.
Consistent performance gains can also be observed via the metrics of ASSD (1.03 $mm$) and 95HD (5.40 $mm$).
Note that the overall improvement achieved by our network on all three evaluation metrics are statistically significant (see the last column in Table~\ref{model_comparison}).

The most recently proposed TransUNet~\cite{chen2021transunet} for abdominal organ segmentation achieved a comparable overall average Dice value of 81.9$\%$, but with significant performance drops comparing to the proposed network in ASSD (3.03 $mm$ vs. 1.03 $mm$) and 95HD (13.25 $mm$ vs. 5.40 $mm$).
The TransUNet employs 2D UNet as backbone and incorporates vision transformers (ViT)~\cite{dosovitskiy2020image} into the encoder.
The performance drop in ASSD and 95HD may because:
(1) The 2D convolution kernels cannot enforce inter-slice consistency, which is critical to volumetric organ segmentation where the locations and shapes of organs are relatively structured.
(2) The patch-based ViT leads to the loss of spatial resolution.
Here, we argue that 3D network architectures are more suitable for volumetric organ segmentation.

Importantly, the proposed network attained statistically significant improvement for the small and complicated structures (e.g., gallbladder, esophagus, stomach, pancreas, and duodenum) over state-of-the-art methods.
The improvement of Dice values are statistically significant in 7/7 comparisons for the stomach and pancreas, 6/7 comparisons for the gallbladder (except the TransUNet) and esophagus (except CascadedVNet), and 5/7 comparisons for the duodenum (except the AttentionUNet and UNet++).
These organs are either small in size (gallbladder and esophagus) or have large shape variability (stomach, pancreas, and duodenum).
The esophagus, stomach and duodenum are from the gastrointestinal tract which are anatomically connected, and they often contains bowel gas which leads to dark areas in CT images.
Better segmentation accuracies for these organs demonstrate the efficacy of the proposed network.
Qualitative comparison between the proposed network and all seven comparisons are presented in Fig.~\ref{fig8}.

For left kidney, the proposed network got higher Dice values than others did, and a marginal 0.2 $mm$ performance drop in ASSD comparing to AttentionUNet (not statistical significant).
The 95HD values had higher variability, thus our value got a larger difference (1.57 $mm$) comparing to AttentionUNet but with no statistical significance.



\subsection{Analytical Ablation Study}

\subsubsection{Effect of Devised Components}
Table~\ref{ablation_study} lists the quantitative results of ablation analyses.
The proposed network had higher Dice values than other comparisons for all organs, and lower average ASSD (1.03 $mm$) and 95HD (5.40 $mm$).
Comparing to the baseline, adding the edge detection sub-network (ED) and edge skip-connections (ESCs) brought 1.2$\%$ performance gain in Dice, and adding the RFP-Head brought 1.4$\%$ improvement.
Eliminating the ED and ESCs yielded a loss in accuracy by all metrics, where the largest loss in Dice value was for the pancreas (from 78.1$\%$ to 76.0$\%$).
Eliminating the RFP-Head yielded less accurate segmentation for most organs, where the performance drop in overlap mainly came from the gallbladder and esophagus (the smallest organs), and duodenum (the least accurate organ).
The results in Table~\ref{ablation_study} demonstrate the merits of the architecture designs, among which the ESCs enable learning a hierarchy of discriminative features, and the RFP-Head computes long-term context for the better segmentation.

\begin{table}[t]
	\centering
	\setlength{\extrarowheight}{1pt}
	\caption{Quantitative comparison of RFP-Head with different number of DAGs. \textbf{Values in boldface denote the best results.} }
	\label{analysis_of_ucg}	
	\begin{tabular}{c|ccc}
		\hline
		\hline
		\# DAGs     & DSC (\%) & ASSD ($mm$) & 95HD ($mm$) \\ \hline
		0 DAGs      & 81.7{\tiny $\pm$10.7}     & 1.22{\tiny $\pm$0.87}      & 6.53{\tiny $\pm$4.27}        \\
		1 DAGs      & 82.3{\tiny $\pm$10.2}     & 1.21{\tiny $\pm$1.22}      & 6.44{\tiny $\pm$4.41}         \\
		2 DAGs      & 83.1{\tiny $\pm$10.0}     & 1.11{\tiny $\pm$1.09}      & 5.76{\tiny $\pm$4.61}          \\ \hline
		4 DAGs      & \textbf{83.4{\tiny $\pm$10.0}}     & \textbf{1.03{\tiny $\pm$0.72}}      & \textbf{5.40{\tiny $\pm$3.61}}         \\ \hline
		\hline
	\end{tabular}
\end{table}

\begin{table}[t]
	\caption{Quantitative comparison of different number of edge skip-connections. \textbf{Values in boldface denote the best results.}}
	\label{analysis_of_escs}
	\centering
	\setlength{\extrarowheight}{1pt}
	\begin{tabular}{c|ccc}
		\hline
		\hline
		\# ESCs        & DSC (\%)      & ASSD (mm)     & 95HD (mm) \\ \hline
		0 ESCs         & 82.1{\tiny $\pm$10.4}          & 1.15{\tiny $\pm$0.97}          & 6.67{\tiny $\pm$4.26}        \\
		1 ESCs         & 82.3{\tiny $\pm$10.7}          & 1.14{\tiny $\pm$0.77}          & 6.26{\tiny $\pm$4.34}         \\ 
		2 ESCs         & 83.1{\tiny $\pm$10.0}          & 1.14{\tiny $\pm$0.79}          & 6.17{\tiny $\pm$4.81}         \\ \hline
		4 ESCs         & \textbf{83.4{\tiny $\pm$10.0}} & \textbf{1.03{\tiny $\pm$0.72}} & \textbf{5.40{\tiny $\pm$3.61}} \\ \hline
		\hline
	\end{tabular}
	
\end{table}

\subsubsection{Different Number of DAGs}
Table~\ref{analysis_of_ucg} shows quantitative comparison of RFP-Head with different number of DAGs.
As seen, only incorporating 1 DAG in our RFP-Head boosted the average Dice by 0.6$\%$, demonstrating the usefulness of context aggregation.
By including 2 DAGs in our RFP-Head, there had a substantial improvement in Dice value (1.4$\%$) comparing to the no DAG variant.
By using 4 DAGs, the results can still be improved but with small rate of increase comparing to the variant with 2 DAGs.
This implies that DAGs sweeping from different directions may somewhat overlap in context aggregation, and incorporating more DAGs would have minimal impact.

\subsubsection{Different Number of Edge Skip-connections}
Table~\ref{analysis_of_escs} shows quantitative comparison of different number of edge skip-connections.
It can be observed that integrating edge features at each decoding blocks achieved the overall best performance.
Simply fusing the edge features at full resolution scale (i.e., 1 ESCs) had minimal impact, but including another connection at 1/2 scale (i.e., 2 ESCs) further improved the performance.
These results demonstrate the effectiveness of enabling multiple edge feature streams to work at different resolutions.

\begin{table}[t]
	\centering
	\setlength{\extrarowheight}{1pt}
	\caption{Quantitative comparison of RFP-Head with four and eight connected neighborhood.}
	\label{number_of_connected_neighbors}
	\begin{tabular}{c|c|c|m{0.8cm}m{0.8cm}m{0.8cm}}
		\hline
		\hline
		\# Neighbors & Param.   & Time  & DSC (\%) & ASSD ($mm$) & 95HD ($mm$) \\ \hline
		UCG(4)       & 6.52 M   & 0.94 s  & 83.4{\tiny $\pm$10.0}  & 1.03{\tiny $\pm$0.72}  & 5.40{\tiny $\pm$3.61}    \\
		UCG(8)       & 6.78 M   & 2.40 s  & 83.4{\tiny $\pm$9.8}   & 0.92{\tiny $\pm$0.79}    & 5.27{\tiny $\pm$3.96}       \\ \hline
		\hline
	\end{tabular}

	\vspace{1mm}
	
	{\raggedright \textit{~~~~~~Time denotes inference time per volume.} \par}

\end{table}

\subsubsection{Neighborhood Connectivity of UCG}
Table~\ref{number_of_connected_neighbors} reports quantitative comparison of RFP-Head with four and eight connected neighborhood.
Introducing UCG(8) brought modest performance gain in ASSD (0.11 $mm$) and 95HD (0.13 $mm$).
However, there was no improvement in Dice value.
This shares the same insight discussed above that the recurrent propagation of feature vectors are somewhat overlapping, thus introducing too many propagation paths would not bring significant improvement.
Note that the UCG(8)'s improvement on ASSD and 95HD was at the expense of memory overhead and computational time.
Therefore, in this paper, we use the RFP-Head with UCG of four connected neighbors in all experimental settings.








\section{Conclusion}
Automatic multi-organ segmentation in CT is an extremely challenging task, due to the issues of low tissue contrast in CT and complicated internal structures in abdomen.
The proposed method provides an effective solution for this significant clinical problem.
The improved performance can be attributed to the edge detector, edge skip-connections and recurrent feature propagation head.
The RFP-Head propagates and harvests local features through directed acylic graphs in an efficient slice-wise manner capturing rich contextual dependencies in respect of spatial arrangement, complementing with the small convolution kernels in FCNs.
The edge detector learns edge priors specifically tuned for semantic segmentation, which are propagated by ESCs to multi-level decoder features to learn a hierarchy of discriminative features. 
We have evaluated the proposed network on two challenging abdominal CT datasets with eight annotated organs.
The proposed network outperforms state-of-the-art networks and achieves statistically significant results for challenging small and complicated structures, including gallbladder, esophagus, stomach, pancreas, and duodenum.




\ifCLASSOPTIONcaptionsoff
  \newpage
\fi



%
\bibliographystyle{IEEEtran.bst}
\bibliography{IEEEabrv,ref}

\begin{thebibliography}{10}
\providecommand{\url}[1]{#1}
\csname url@samestyle\endcsname
\providecommand{\newblock}{\relax}
\providecommand{\bibinfo}[2]{#2}
\providecommand{\BIBentrySTDinterwordspacing}{\spaceskip=0pt\relax}
\providecommand{\BIBentryALTinterwordstretchfactor}{4}
\providecommand{\BIBentryALTinterwordspacing}{\spaceskip=\fontdimen2\font plus
\BIBentryALTinterwordstretchfactor\fontdimen3\font minus
  \fontdimen4\font\relax}
\providecommand{\BIBforeignlanguage}[2]{{%
\expandafter\ifx\csname l@#1\endcsname\relax
\typeout{** WARNING: IEEEtran.bst: No hyphenation pattern has been}%
\typeout{** loaded for the language `#1'. Using the pattern for}%
\typeout{** the default language instead.}%
\else
\language=\csname l@#1\endcsname
\fi
#2}}
\providecommand{\BIBdecl}{\relax}
\BIBdecl

\bibitem{van2011computer}
B.~Van~Ginneken, C.~M. Schaefer-Prokop, and M.~Prokop, ``Computer-aided
  diagnosis: how to move from the laboratory to the clinic,'' \emph{Radiology},
  vol. 261, no.~3, pp. 719--732, 2011.

\bibitem{tang2019clinically}
H.~Tang, X.~Chen, Y.~Liu, Z.~Lu, J.~You, M.~Yang, S.~Yao, G.~Zhao, Y.~Xu,
  T.~Chen \emph{et~al.}, ``Clinically applicable deep learning framework for
  organs at risk delineation in ct images,'' \emph{Nature Machine
  Intelligence}, vol.~1, no.~10, pp. 480--491, 2019.

\bibitem{graves2007multi}
A.~Graves, S.~Fern{\'a}ndez, and J.~Schmidhuber, ``Multi-dimensional recurrent
  neural networks,'' in \emph{International conference on artificial neural
  networks}.\hskip 1em plus 0.5em minus 0.4em\relax Springer, 2007, pp.
  549--558.

\bibitem{zhang2010automatic}
X.~Zhang, J.~Tian, K.~Deng, Y.~Wu, and X.~Li, ``Automatic liver segmentation
  using a statistical shape model with optimal surface detection,'' \emph{IEEE
  Transactions on Biomedical Engineering}, vol.~57, no.~10, pp. 2622--2626,
  2010.

\bibitem{cerrolaza2015automatic}
J.~J. Cerrolaza, M.~Reyes, R.~M. Summers, M.~{\'A}. Gonz{\'a}lez-Ballester, and
  M.~G. Linguraru, ``Automatic multi-resolution shape modeling of multi-organ
  structures,'' \emph{Med. Image Anal.}, vol.~25, no.~1, pp. 11--21, 2015.

\bibitem{tong2015discriminative}
T.~Tong, R.~Wolz, Z.~Wang, Q.~Gao, K.~Misawa, M.~Fujiwara, K.~Mori, J.~V.
  Hajnal, and D.~Rueckert, ``Discriminative dictionary learning for abdominal
  multi-organ segmentation,'' \emph{Med. Image Anal.}, vol.~23, no.~1, pp.
  92--104, 2015.

\bibitem{xu2015efficient}
Z.~Xu, R.~P. Burke, C.~P. Lee, R.~B. Baucom, B.~K. Poulose, R.~G. Abramson, and
  B.~A. Landman, ``Efficient multi-atlas abdominal segmentation on clinically
  acquired ct with simple context learning,'' \emph{Medical image analysis},
  vol.~24, no.~1, pp. 18--27, 2015.

\bibitem{shimizu2007segmentation}
A.~Shimizu, R.~Ohno, T.~Ikegami, H.~Kobatake, S.~Nawano, and D.~Smutek,
  ``Segmentation of multiple organs in non-contrast 3d abdominal ct images,''
  \emph{International journal of computer assisted radiology and surgery},
  vol.~2, no.~3, pp. 135--142, 2007.

\bibitem{long2015fully}
J.~Long, E.~Shelhamer, and T.~Darrell, ``Fully convolutional networks for
  semantic segmentation,'' in \emph{Proceedings of the IEEE conference on
  computer vision and pattern recognition}, 2015, pp. 3431--3440.

\bibitem{ronneberger2015u}
O.~Ronneberger, P.~Fischer, and T.~Brox, ``U-net: Convolutional networks for
  biomedical image segmentation,'' in \emph{International Conference on Medical
  image computing and computer-assisted intervention}.\hskip 1em plus 0.5em
  minus 0.4em\relax Springer, 2015, pp. 234--241.

\bibitem{cciccek20163d}
{\"O}.~{\c{C}}i{\c{c}}ek, A.~Abdulkadir, S.~S. Lienkamp, T.~Brox, and
  O.~Ronneberger, ``3d u-net: learning dense volumetric segmentation from
  sparse annotation,'' in \emph{MICCAI 2016}.\hskip 1em plus 0.5em minus
  0.4em\relax Springer, 2016, pp. 424--432.

\bibitem{milletari2016v}
F.~Milletari, N.~Navab, and S.-A. Ahmadi, ``V-net: Fully convolutional neural
  networks for volumetric medical image segmentation,'' in \emph{2016 fourth
  international conference on 3D vision (3DV)}.\hskip 1em plus 0.5em minus
  0.4em\relax IEEE, 2016, pp. 565--571.

\bibitem{oktay2018attention}
O.~Oktay, J.~Schlemper, L.~L. Folgoc, M.~Lee, M.~Heinrich, K.~Misawa, K.~Mori,
  S.~McDonagh, N.~Y. Hammerla, B.~Kainz \emph{et~al.}, ``Attention u-net:
  Learning where to look for the pancreas,'' \emph{arXiv preprint
  arXiv:1804.03999}, 2018.

\bibitem{zhou2019unet++}
Z.~Zhou, M.~M.~R. Siddiquee, N.~Tajbakhsh, and J.~Liang, ``Unet++: Redesigning
  skip connections to exploit multiscale features in image segmentation,''
  \emph{IEEE transactions on medical imaging}, vol.~39, no.~6, pp. 1856--1867,
  2019.

\bibitem{li2018h}
X.~Li, H.~Chen, X.~Qi, Q.~Dou, C.-W. Fu, and P.-A. Heng, ``H-denseunet: hybrid
  densely connected unet for liver and tumor segmentation from ct volumes,''
  \emph{IEEE transactions on medical imaging}, vol.~37, no.~12, pp. 2663--2674,
  2018.

\bibitem{seo2019modified}
H.~Seo, C.~Huang, M.~Bassenne, R.~Xiao, and L.~Xing, ``Modified u-net (mu-net)
  with incorporation of object-dependent high level features for improved liver
  and liver-tumor segmentation in ct images,'' \emph{IEEE transactions on
  medical imaging}, vol.~39, no.~5, pp. 1316--1325, 2019.

\bibitem{roth2018spatial}
H.~R. Roth, L.~Lu, N.~Lay, A.~P. Harrison, A.~Farag, A.~Sohn, and R.~M.
  Summers, ``Spatial aggregation of holistically-nested convolutional neural
  networks for automated pancreas localization and segmentation,''
  \emph{Medical image analysis}, vol.~45, pp. 94--107, 2018.

\bibitem{xue2019cascaded}
J.~Xue, K.~He, D.~Nie, E.~Adeli, Z.~Shi, S.-W. Lee, Y.~Zheng, X.~Liu, D.~Li,
  and D.~Shen, ``Cascaded multitask 3-d fully convolutional networks for
  pancreas segmentation,'' \emph{IEEE Transactions on Cybernetics}, 2019.

\bibitem{gibson2017towards}
E.~Gibson, F.~Giganti, Y.~Hu, E.~Bonmati, S.~Bandula, K.~Gurusamy, B.~R.
  Davidson, S.~P. Pereira, M.~J. Clarkson, and D.~C. Barratt, ``Towards
  image-guided pancreas and biliary endoscopy: Automatic multi-organ
  segmentation on abdominal ct with dense dilated networks,'' in
  \emph{International Conference on Medical Image Computing and
  Computer-Assisted Intervention}.\hskip 1em plus 0.5em minus 0.4em\relax
  Springer, 2017, pp. 728--736.

\bibitem{gibson2018automatic}
E.~Gibson, F.~Giganti, Y.~Hu, E.~Bonmati, S.~Bandula, K.~Gurusamy, B.~Davidson,
  S.~P. Pereira, M.~J. Clarkson, and D.~C. Barratt, ``Automatic multi-organ
  segmentation on abdominal ct with dense v-networks,'' \emph{IEEE Trans. Med.
  Imaging}, vol.~37, no.~8, pp. 1822--1834, 2018.

\bibitem{huang2017densely}
G.~Huang, Z.~Liu, L.~Van Der~Maaten, and K.~Q. Weinberger, ``Densely connected
  convolutional networks,'' in \emph{Proceedings of the IEEE conference on
  computer vision and pattern recognition}, 2017, pp. 4700--4708.

\bibitem{roth2018multi}
H.~R. Roth, C.~Shen, H.~Oda, T.~Sugino, M.~Oda, Y.~Hayashi, K.~Misawa, and
  K.~Mori, ``A multi-scale pyramid of 3d fully convolutional networks for
  abdominal multi-organ segmentation,'' in \emph{International conference on
  medical image computing and computer-assisted intervention}.\hskip 1em plus
  0.5em minus 0.4em\relax Springer, 2018, pp. 417--425.

\bibitem{roth2018application}
H.~R. Roth, H.~Oda, X.~Zhou, N.~Shimizu, Y.~Yang, Y.~Hayashi, M.~Oda,
  M.~Fujiwara, K.~Misawa, and K.~Mori, ``An application of cascaded 3d fully
  convolutional networks for medical image segmentation,'' \emph{Computerized
  Medical Imaging and Graphics}, vol.~66, pp. 90--99, 2018.

\bibitem{wang2019abdominal}
Y.~Wang, Y.~Zhou, W.~Shen, S.~Park, E.~K. Fishman, and A.~L. Yuille,
  ``Abdominal multi-organ segmentation with organ-attention networks and
  statistical fusion,'' \emph{Medical image analysis}, vol.~55, pp. 88--102,
  2019.

\bibitem{zhang2020block}
L.~Zhang, J.~Zhang, P.~Shen, G.~Zhu, P.~Li, X.~Lu, H.~Zhang, S.~A. Shah, and
  M.~Bennamoun, ``Block level skip connections across cascaded v-net for
  multi-organ segmentation,'' \emph{IEEE Trans. Med. Imaging}, vol.~39, no.~9,
  pp. 2782--2793, 2020.

\bibitem{heinrich2019obelisk}
M.~P. Heinrich, O.~Oktay, and N.~Bouteldja, ``Obelisk-net: Fewer layers to
  solve 3d multi-organ segmentation with sparse deformable convolutions,''
  \emph{Medical image analysis}, vol.~54, pp. 1--9, 2019.

\bibitem{kamnitsas2017efficient}
K.~Kamnitsas, C.~Ledig, V.~F. Newcombe, J.~P. Simpson, A.~D. Kane, D.~K. Menon,
  D.~Rueckert, and B.~Glocker, ``Efficient multi-scale 3d cnn with fully
  connected crf for accurate brain lesion segmentation,'' \emph{Medical image
  analysis}, vol.~36, pp. 61--78, 2017.

\bibitem{krahenbuhl2011efficient}
P.~Kr{\"a}henb{\"u}hl and V.~Koltun, ``Efficient inference in fully connected
  crfs with gaussian edge potentials,'' \emph{Advances in neural information
  processing systems}, vol.~24, pp. 109--117, 2011.

\bibitem{christ2016automatic}
P.~F. Christ, M.~E.~A. Elshaer, F.~Ettlinger, S.~Tatavarty, M.~Bickel,
  P.~Bilic, M.~Rempfler, M.~Armbruster, F.~Hofmann, M.~D’Anastasi
  \emph{et~al.}, ``Automatic liver and lesion segmentation in ct using cascaded
  fully convolutional neural networks and 3d conditional random fields,'' in
  \emph{International Conference on Medical Image Computing and
  Computer-Assisted Intervention}.\hskip 1em plus 0.5em minus 0.4em\relax
  Springer, 2016, pp. 415--423.

\bibitem{alansary2016fast}
A.~Alansary, K.~Kamnitsas, A.~Davidson, R.~Khlebnikov, M.~Rajchl,
  C.~Malamateniou, M.~Rutherford, J.~V. Hajnal, B.~Glocker, D.~Rueckert
  \emph{et~al.}, ``Fast fully automatic segmentation of the human placenta from
  motion corrupted mri,'' in \emph{International conference on medical image
  computing and computer-assisted intervention}.\hskip 1em plus 0.5em minus
  0.4em\relax Springer, 2016, pp. 589--597.

\bibitem{poudel2016recurrent}
R.~P. Poudel, P.~Lamata, and G.~Montana, ``Recurrent fully convolutional neural
  networks for multi-slice mri cardiac segmentation,'' in \emph{Reconstruction,
  segmentation, and analysis of medical images}.\hskip 1em plus 0.5em minus
  0.4em\relax Springer, 2016, pp. 83--94.

\bibitem{chen2016combining}
J.~Chen, L.~Yang, Y.~Zhang, M.~S. Alber, and D.~Z. Chen, ``Combining fully
  convolutional and recurrent neural networks for 3d biomedical image
  segmentation,'' in \emph{NIPS}, 2016.

\bibitem{xie2016spatial}
Y.~Xie, Z.~Zhang, M.~Sapkota, and L.~Yang, ``Spatial clockwork recurrent neural
  network for muscle perimysium segmentation,'' in \emph{International
  Conference on Medical Image Computing and Computer-Assisted
  Intervention}.\hskip 1em plus 0.5em minus 0.4em\relax Springer, 2016, pp.
  185--193.

\bibitem{shuai2017scene}
B.~Shuai, Z.~Zuo, B.~Wang, and G.~Wang, ``Scene segmentation with dag-recurrent
  neural networks,'' \emph{IEEE Trans. Pattern Anal. Mach. Intell.}, vol.~40,
  no.~6, pp. 1480--1493, 2017.

\bibitem{stollenga2015parallel}
M.~F. Stollenga, W.~Byeon, M.~Liwicki, and J.~Schmidhuber, ``Parallel
  multi-dimensional lstm, with application to fast biomedical volumetric image
  segmentation,'' \emph{Advances in neural information processing systems},
  vol.~28, pp. 2998--3006, 2015.

\bibitem{andermatt2016multi}
S.~Andermatt, S.~Pezold, and P.~Cattin, ``Multi-dimensional gated recurrent
  units for the segmentation of biomedical 3d-data,'' in \emph{Deep learning
  and data labeling for medical applications}.\hskip 1em plus 0.5em minus
  0.4em\relax Springer, 2016, pp. 142--151.

\bibitem{anas2017clinical}
E.~M.~A. Anas, S.~Nouranian, S.~S. Mahdavi, I.~Spadinger, W.~J. Morris, S.~E.
  Salcudean, P.~Mousavi, and P.~Abolmaesumi, ``Clinical target-volume
  delineation in prostate brachytherapy using residual neural networks,'' in
  \emph{International Conference on Medical Image Computing and
  Computer-Assisted Intervention}.\hskip 1em plus 0.5em minus 0.4em\relax
  Springer, 2017, pp. 365--373.

\bibitem{karimi2019reducing}
D.~Karimi and S.~E. Salcudean, ``Reducing the hausdorff distance in medical
  image segmentation with convolutional neural networks,'' \emph{IEEE
  Transactions on medical imaging}, vol.~39, no.~2, pp. 499--513, 2019.

\bibitem{ma2020learning}
J.~Ma, J.~He, and X.~Yang, ``Learning geodesic active contours for embedding
  object global information in segmentation cnns,'' \emph{IEEE Transactions on
  Medical Imaging}, 2020.

\bibitem{fan2020inf}
D.-P. Fan, T.~Zhou, G.-P. Ji, Y.~Zhou, G.~Chen, H.~Fu, J.~Shen, and L.~Shao,
  ``Inf-net: Automatic covid-19 lung infection segmentation from ct images,''
  \emph{IEEE Transactions on Medical Imaging}, vol.~39, no.~8, pp. 2626--2637,
  2020.

\bibitem{zhang2019net}
Z.~Zhang, H.~Fu, H.~Dai, J.~Shen, Y.~Pang, and L.~Shao, ``Et-net: A generic
  edge-attention guidance network for medical image segmentation,'' in
  \emph{International Conference on Medical Image Computing and
  Computer-Assisted Intervention}.\hskip 1em plus 0.5em minus 0.4em\relax
  Springer, 2019, pp. 442--450.

\bibitem{zhou2019high}
S.~Zhou, D.~Nie, E.~Adeli, J.~Yin, J.~Lian, and D.~Shen, ``High-resolution
  encoder--decoder networks for low-contrast medical image segmentation,''
  \emph{IEEE Transactions on Image Processing}, vol.~29, pp. 461--475, 2019.

\bibitem{lee2015deeply}
C.-Y. Lee, S.~Xie, P.~Gallagher, Z.~Zhang, and Z.~Tu, ``Deeply-supervised
  nets,'' in \emph{Artificial intelligence and statistics}.\hskip 1em plus
  0.5em minus 0.4em\relax PMLR, 2015, pp. 562--570.

\bibitem{chen2016semantic}
L.-C. Chen, J.~T. Barron, G.~Papandreou, K.~Murphy, and A.~L. Yuille,
  ``Semantic image segmentation with task-specific edge detection using cnns
  and a discriminatively trained domain transform,'' in \emph{Proceedings of
  the IEEE conference on computer vision and pattern recognition}, 2016, pp.
  4545--4554.

\bibitem{visin2016reseg}
F.~Visin, M.~Ciccone, A.~Romero, K.~Kastner, K.~Cho, Y.~Bengio, M.~Matteucci,
  and A.~Courville, ``Reseg: A recurrent neural network-based model for
  semantic segmentation,'' in \emph{Proceedings of the IEEE Conference on
  Computer Vision and Pattern Recognition Workshops}, 2016, pp. 41--48.

\bibitem{canny1986computational}
J.~Canny, ``A computational approach to edge detection,'' \emph{IEEE
  Transactions on pattern analysis and machine intelligence}, no.~6, pp.
  679--698, 1986.

\bibitem{roth2015deeporgan}
H.~R. Roth, L.~Lu, A.~Farag, H.-C. Shin, J.~Liu, E.~B. Turkbey, and R.~M.
  Summers, ``Deeporgan: Multi-level deep convolutional networks for automated
  pancreas segmentation,'' in \emph{MICCAI 2015}.\hskip 1em plus 0.5em minus
  0.4em\relax Springer, 2015, pp. 556--564.

\bibitem{wang2019deep}
Y.~Wang, H.~Dou, X.~Hu, L.~Zhu, X.~Yang, M.~Xu, J.~Qin, P.-A. Heng, T.~Wang,
  and D.~Ni, ``Deep attentive features for prostate segmentation in 3d
  transrectal ultrasound,'' \emph{IEEE Transactions on Medical Imaging},
  vol.~38, no.~12, pp. 2768--2778, 2019.

\bibitem{chen2021transunet}
J.~Chen, Y.~Lu, Q.~Yu, X.~Luo, E.~Adeli, Y.~Wang, L.~Lu, A.~L. Yuille, and
  Y.~Zhou, ``Transunet: Transformers make strong encoders for medical image
  segmentation,'' \emph{arXiv preprint arXiv:2102.04306}, 2021.

\bibitem{he2016deep}
K.~He, X.~Zhang, S.~Ren, and J.~Sun, ``Deep residual learning for image
  recognition,'' in \emph{CVPR}, 2016, pp. 770--778.

\bibitem{vaswani2017attention}
A.~Vaswani, N.~Shazeer, N.~Parmar, J.~Uszkoreit, L.~Jones, A.~N. Gomez,
  {\L}.~Kaiser, and I.~Polosukhin, ``Attention is all you need,'' in
  \emph{Advances in neural information processing systems}, 2017, pp.
  5998--6008.

\bibitem{dosovitskiy2020image}
A.~Dosovitskiy, L.~Beyer, A.~Kolesnikov, D.~Weissenborn, X.~Zhai,
  T.~Unterthiner, M.~Dehghani, M.~Minderer, G.~Heigold, S.~Gelly \emph{et~al.},
  ``An image is worth 16x16 words: Transformers for image recognition at
  scale,'' \emph{arXiv preprint arXiv:2010.11929}, 2020.

\end{thebibliography}

%








\end{document}